\documentclass[lettersize,journal]{IEEEtran}
\usepackage{amsmath,amsfonts}
\usepackage{algorithmic}
\usepackage{algorithm}
\usepackage{array}
\usepackage[caption=false,font=normalsize,labelfont=sf,textfont=sf]{subfig}
\usepackage{textcomp}
\usepackage{stfloats}
\usepackage{url}
\usepackage{verbatim}
\usepackage{graphicx}
\usepackage{cite}
\usepackage[export]{adjustbox}
\usepackage{xcolor}
\newif\ifdraft
\drafttrue

\ifdraft
\newcommand{\mycolor}[1]{#1}
\newcommand{\note}[1]{\textit{\textcolor{magenta}{#1}}}
\newcommand{\figDir}{low}
\else
\newcommand{\mycolor}[1]{#1}
\newcommand{\note}[1]{}
\newcommand{\figDir}{high}
\fi

\DeclareMathOperator*{\argmin}{arg\,min}
\DeclareMathOperator*{\argmax}{arg\,max}
\newcommand{\norm}[1]{\left\lVert#1\right\rVert}
\newcommand{\realNumber}{\mathbb{R}}

\newcommand{\generalizedPos}{\mathbf{q}}
\newcommand{\generalizedVel}{\mathbf{\dot{q}}}
\newcommand{\generalizedForce}{\mathbf{f}}
\newcommand{\generalizedForceExt}{\mathbf{f}_{\mathrm{ext}}}
\newcommand{\generalizedMass}{\mathbf{M}}
\newcommand{\vertexPos}{\mathbf{x}}

\newcommand{\edgeAngle}{\boldsymbol{\theta}}

\newcommand{\dt}{\Delta t}
\newcommand{\radius}{r}

\newcommand{\length}{\mathbf{l}}
\newcommand{\tangent}{\mathbf{t}}
\newcommand{\curvature}{\boldsymbol{\kappa}}
\newcommand{\twist}{\mathbf{m}}

\newcommand{\restLength}{\mathbf{\bar{l}}}
\newcommand{\restCurvature}{\boldsymbol{\bar{\kappa}}}
\newcommand{\restTwist}{\mathbf{\bar{m}}}

\newcommand{\penaltyCoef}{\rho}

\newcommand{\stretchCoef}{\boldsymbol{\alpha}}
\newcommand{\bendCoef}{\boldsymbol{\beta}}
\newcommand{\twistCoef}{\boldsymbol{\gamma}}

\newcommand{\Jacobian}{\mathbf{J}}
\newcommand{\Imat}{\mathbf{I}}

\newcommand{\evec}{\mathbf{e}}

\newcommand{\parameter}{\mathbf{p}}
\newcommand{\parameterMin}{\mathbf{p}_{\mathrm{min}}}
\newcommand{\parameterMax}{\mathbf{p}_{\mathrm{max}}}
\newcommand{\constraint}{\mathbf{c}}
\newcommand{\multiplier}{\boldsymbol{\lambda}}
\newcommand{\iterIndex}{k}

\newcommand{\yvec}{\mathbf{y}}

\newcommand{\Amat}{\mathbf{A}}
\newcommand{\Lmat}{\mathbf{L}}
\newcommand{\xvec}{\mathbf{x}}
\newcommand{\bvec}{\mathbf{b}}
\newcommand{\rvec}{\mathbf{r}}
\newcommand{\zvec}{\mathbf{z}}
\newcommand{\avec}{\mathbf{a}}
\newcommand{\Smat}{\mathbf{S}}
\newcommand{\Dmat}{\mathbf{D}}
\newcommand{\tLmat}{\tilde{\Lmat}}
\newcommand{\tDmat}{\tilde{\Dmat}}
\newcommand{\hLmat}{\hat{\Lmat}}

\newcommand{\wvec}{\mathbf{w}}

\newcommand{\Wmat}{\mathbf{W}}

\newcommand{\curvRange}{\mu}
\newcommand{\stiffnessScale}{s}
\newcommand{\stretchCoefFinal}{\mathbf{c}_{\mathrm{st}}}
\newcommand{\bendCoefFinal}{\mathbf{c}_{\mathrm{be}}}
\newcommand{\twistCoefFinal}{\mathbf{c}_{\mathrm{tw}}}
\newcommand{\stretchCoefFinali}{\mathbf{c}_{\mathrm{st}, i}}
\newcommand{\bendCoefFinali}{\mathbf{c}_{\mathrm{be}, i}}
\newcommand{\twistCoefFinali}{\mathbf{c}_{\mathrm{tw}, i}}

\begin{document}


\title{%
Optimizing Parameters for Static Equilibrium of Discrete Elastic Rods with Active-Set Cholesky
}


\author
{
Tetsuya Takahashi and Christopher Batty
\thanks{Tetsuya Takahashi is with Tencent America, USA; Christopher Batty is with University of Waterloo, Canada}
}




\maketitle

\begin{abstract}
We propose a parameter optimization method for achieving static equilibrium of discrete elastic rods. Our method simultaneously optimizes material stiffness and rest shape parameters under box constraints to exactly enforce zero net forces while avoiding stability issues and violations of physical laws. For efficiency, we split our constrained optimization problem into primal and dual subproblems via the augmented Lagrangian method, while handling the dual maximization subproblem via simple vector updates. To efficiently solve the box-constrained primal minimization subproblem, we propose a new active-set Cholesky preconditioner for variants of conjugate gradient solvers with active sets. Our method surpasses prior work in generality, robustness, and speed.
\end{abstract}






\section{Introduction}
When modeling deformable objects or fabricating elastic materials, sagging under external forces, such as gravity and applied loads, can ruin specific shapes carefully designed by artists and designers or captured from real world counterparts. This sagging problem arises because the external forces are often implicitly considered during modeling but simulators typically neglect it during initialization \cite{Twigg:2011:OSS:2019406.2019437,Perez2015rod,Hsu2022sag,Hsu2023sag,Takahashi2025rest}. To prevent this issue, we aim to achieve static equilibrium of the objects preserving their original shapes at initialization. While stiffening materials can reduce the sagging, it alters the dynamic response of the elastic material. Alternatively, modifying rest shape parameters can achieve static equilibrium, though substantial rest shape changes can introduce stability problems or increase the likelihood of undesirable rest configurations \cite{Takahashi2025rest}.

In this paper, we focus on elastic objects with one-dimensional structures (e.g., hair, cables, strands, etc.). We employ discrete elastic rods (DER) \cite{Bergou2008,Bergou:2010:DVT:1778765.1778853} for strand simulation due to their generality, flexibility, and efficiency, as adopted in various applications with forward simulations \cite{Schweickart2017,Fei2019,Lesser2022,Daviet2023hair} and inverse problems \cite{Perez2015rod, Panetta2019,Ren2022umbrella,Dandy2024rods}. 

\begin{figure*}
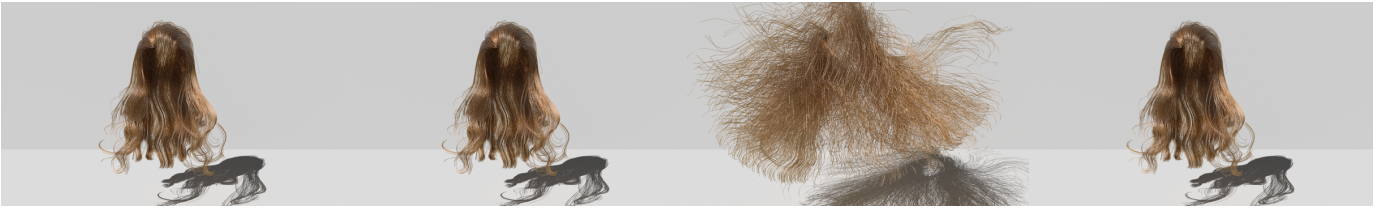
 
\centering
    \adjincludegraphics[trim={{0.05\width} {0.0\height} {0.0\width} {0.0\height}}, clip, width=0.25\linewidth]
    {figures/\figDir/der_0001.png}%
    \adjincludegraphics[trim={{0.05\width} {0.0\height} {0.0\width} {0.0\height}}, clip, width=0.25\linewidth]
    {figures/\figDir/der_0181.png}%
    \adjincludegraphics[trim={{0.05\width} {0.0\height} {0.0\width} {0.0\height}}, clip, width=0.25\linewidth]
    {figures/\figDir/der_0225.png}%
    \adjincludegraphics[trim={{0.05\width} {0.0\height} {0.0\width} {0.0\height}}, clip, width=0.25\linewidth]
    {figures/\figDir/der_1200.png}%
\caption{
Our method optimizes material stiffness and rest shape parameters to achieve static equilibrium for DER-based strands. It preserves the original hairstyle without sagging, demonstrates natural dynamic strand behaviors in response to prescribed motions of the root vertices, and eventually restores the strands to their initial configuration. Our parameter optimization required 713 seconds for 3.2k strands; forward simulation without collision handling took 25 seconds per frame.
\vspace{-3mm}
}
\label{fig:teaser} 
\end{figure*}

To address the sagging problem, we propose a new parameter optimization method that is guaranteed (at solver convergence) to achieve static equilibrium of DER-based strands. Our method simultaneously optimizes rest shape and material stiffness parameters while minimizing changes in the parameters and satisfying their corresponding box constraints; the latter are imposed to avoid violating physical laws or introducing stability issues in forward simulation. We formulate our problem as a constrained minimization and decompose it into primal and dual subproblems via the augmented Lagrangian method (ALM) \cite{NoceWrig06}, addressing the primal minimization subproblem with a Newton-type optimizer and the dual maximization subproblem via vector updates. To efficiently and accurately handle the primal subproblem, we take the box constraints into account and solve the symmetric positive definite (SPD) inner systems using a variant of conjugate gradient (CG) with active sets, modified proportioning with reduced gradient projections (MPRGP) \cite{Dostal2005}. Additionally, we propose a new preconditioner based on a Cholesky-based direct solver with active sets to accelerate the convergence of MPRGP. Figure \ref{fig:teaser} shows our method in action.


\section{Related Work}


\subsection{Elastic Rod Simulation}
To capture bending and twisting of one-dimensional elastic strand structures, Pai \cite{Pai2002cosserat} introduced the Cosserat theory into graphics, which was later extended to dynamical systems \cite{Spillmann2007rod}. The Cosserat theory has since been adopted within position-based dynamics \cite{Umetani2014rod} and further extended with quaternions \cite{Kugelstadt2016rod,Hsu2023sag}, volume constraints \cite{Angles2019rods},
and projective dynamics \cite{Soler2018rod}. Instead of evolving edge rotations, Shi et al.~\cite{Shi2023curls} proposed using a volume-like torsion energy to capture twisting effects. Curvature-based approaches were also presented to capture elastic rod dynamics with a smaller number of degrees of freedom (DOF) \cite{Bertails2006hair,Casati2013clothoids}. From the various methods for simulating elastic strands, we selected DER \cite{Bergou2008,Bergou:2010:DVT:1778765.1778853} because it can handle general bending and twisting cases with few DOFs.


\subsection{Sag-Free Simulation}
To achieve sag-free simulations (or equivalently, static equilibrium), one common approach is to enforce zero net forces in the dynamical system. Hadap \cite{Hadap2006strands} proposed using inverse dynamics to solve nonlinear force equations for reduced multibody systems, although such methods may not necessarily be applicable to general maximal coordinate systems (e.g., finite-element-method-based deformables and DER). To solve nonlinear force equations, the asymptotic numerical method (ANM) was employed, achieving faster convergence compared to Newton-type optimizers \cite{Chen2014asymptotic,Jia2021SANM}, although it remains unclear how to incorporate box constraints, which help prevent stability problems and violations of physical laws. Curvature-based solvers have also been employed, since they need fewer DOFs to represent elastic rod dynamics. Derouet-Jourdan et al. \cite{Derouet-Jourdan2010} presented a method for achieving static equilibrium of curvature-based elastic strands \cite{Bertails2006hair} and extended their work with frictional contacts \cite{Derouet-Jourdan:2013:IDH:2508363.2508398}. The curvature-based formulations have extensively been used for inverse problems to model, design, and fabricate elastic rods \cite{bertailsdescoubes:hal-01827887,Charrondiere2020,Hafner2021curves,Hafner2023rod}.

To achieve static equilibrium in more general contexts, Twigg and Ka\v{c}i\'{c}-Alesi\'{c} proposed minimizing the force norm by optimizing rest shape parameters in mass-spring systems consisting of one-dimensional elements \cite{Twigg:2011:OSS:2019406.2019437}. Takahashi and Batty extended their approach to DER \cite{Takahashi2025rest} and proposed optimizing the rest length, rest curvature, and rest twist parameters while penalizing the force norm in the kinetic energy metric. However, this approach would fail to achieve static equilibrium because fully optimized rest shape parameters alone could still be insufficient to achieve zero net forces. Sag-free DER simulations have been attempted in other previous work as well \cite{Aubry2015,Lesser2022}, albeit without offering formulation and algorithmic specifics. Rest shape optimization has also been applied to two-dimensional sheet-like structures \cite{Skouras2012design} and three-dimensional deformable volumes \cite{Wang:2015:DCM:2809654.2766911,Mukherjee2018shape}. To enforce zero-net-force constraints, adjoint methods have been extensively used \cite{Perez2015rod,Ly2018shell,Panetta2019}. While this approach can be efficient and \mycolor{can} ensure stable local minima by examining the Hessian of the DER objectives, handling the full Hessian due to differentiation of nonlinear force constraints can be complicated \cite{Panetta2019,Fei2019}. Given that DER Hessian treatments are the subject of ongoing study \cite{Panetta2019,Shi2023curls}, we formulate our problem without explicitly incorporating the Hessian into the parameter optimization. While finite differences could bypass this complexity \cite{Choi2024}, they are known to be significantly slower than optimization with analytical gradients \cite{NoceWrig06}.

As an alternative, Hsu et al. \cite{Hsu2022sag,Hsu2023sag} proposed a global-local initialization method that first computes global forces to achieve static equilibrium and then adjusts local elements to generate such forces. However, this approach would require excessive stiffening of materials and is applicable only to specific simulation approaches that can locally define forces without coupling neighboring elements. As such, the global-local initialization is not applicable to DER \cite{Takahashi2025rest}.


\subsection{Preconditioning with Active Sets}
Preconditioners have been extensively used to reduce the number of necessary iterations for iterative solvers, such as CG. Preconditioners are also applicable and effective for variants of CG with an active-set method, e.g., MPRGP (which updates active sets in each iteration to find their optimal sets and solution) \cite{Dostal2005}, by taking active sets into account during the preconditioning.

While Narain et al. \cite{Narain:2010:FGM:1882261.1866195} used modified incomplete Cholesky (MIC) preconditioning \cite{Bridson2007course} for MPRGP, the derivations and rationale behind their approach are not explained, and it is not necessarily clear from their publicly available source code how to adapt it to support variants of Cholesky solvers (e.g., LDLT-type decomposition with $2\times2$ diagonal blocks \cite{Greif2017ldlt}). Recently, Takahashi and Batty \cite{Takahashi2023} proposed a smoothed aggregation algebraic multigrid (SAAMG) preconditioner for MPRGP, using an SPAI-0 smoother \cite{Broker2001} with active sets (which can easily be modified to a diagonal, weighted Jacobi, or symmetric Gauss-Seidel (SGS) smoother/preconditioner). However, their focus is on systems with M-matrices, and thus the effectiveness of their preconditioning on non-M-matrices, such as those in our formulation, remains unverified.



\section{Contributions}
\label{sec:contributions}
We propose optimizing both the material stiffness and rest shape parameters simultaneously while enforcing zero net forces as a hard constraint via ALM to ensure the static equilibrium of DER-based strands at solver convergence. To reduce the optimization cost, we propose two techniques. First, we arrange the optimization parameters in an interleaved way to yield a banded system by leveraging the one-dimensional strand structure. This arrangement eliminates the need for matrix reordering in Cholesky-based solvers. Second, given the overlapping force spaces with 4D curvatures in the bending formulation \cite{Bergou2008,Fei2019}, we consider only two rest curvature variables per vertex in the optimization, reducing the DOF count and improving efficiency. Moreover, instead of using penalty-based methods \cite{Takahashi2025rest}, we employ an active-set-based iterative solver, MPRGP \cite{Dostal2005}, to enforce the box constraints precisely. We also propose a new active-set Cholesky preconditioner to accelerate the convergence of MPRGP. \mycolor{Because of these improvements, our method can more robustly and efficiently achieve static equilibrium of DER-based strands.}

\section{Discrete Elastic Rods Preliminaries}
\label{sec:DER}
For completeness, we briefly summarize key details of the DER formulation \cite{Bergou2008,Bergou:2010:DVT:1778765.1778853} before explaining our parameter optimization method (Sec. \ref{sec:parameter_optimization}) and active-set Cholesky preconditioner (Sec. \ref{sec:active_set_cholesky}). In our framework, we exclude contacts (as our goal is to achieve static equilibrium which introduces no contacts if none exist at initialization) and thus process each strand independently, focusing on a single strand below.

Given a strand discretized into $N$ vertices with positions $\vertexPos = [\vertexPos_0^T, \ldots, \vertexPos_{N-1}^T]^T \in \realNumber^{3N}$ and $N-1$ edges with angles $\edgeAngle = [\theta_0, \ldots, \theta_{N-2}]^T \in \realNumber^{N-1}$, the generalized positions can be defined by interleaving vertex and edge variables as $\generalizedPos = [\vertexPos_0^T, \theta_0, \ldots, \theta_{N-2}, \vertexPos_{N-1}^T]^T \in \realNumber^{4N-1}$. This arrangement leads to a banded Hessian due to the locally defined DER objectives and time-parallel transport \cite{Bergou:2010:DVT:1778765.1778853}. For simplicity, we assume a strand with a perfectly circular cross-section and uniform radius $\radius$. The root end of the strand is minimally clamped, fixing $\vertexPos_0, \edgeAngle_0$, and $\vertexPos_1$, which yields $4N-8$ active DOFs \cite{Takahashi2025rest}. For more details, we refer to a book on DER \cite{jawed2018primer}.

We define a minimization problem for a forward simulation step with a DER objective $E(\generalizedPos)$ \cite{Takahashi2025rest} as $\generalizedPos = \argmin_{\generalizedPos} E(\generalizedPos)$, where $E(\generalizedPos) = 
E_{\mathrm{in}}(\generalizedPos) +
E_{\mathrm{st}}(\generalizedPos) +
E_{\mathrm{be}}(\generalizedPos) +
E_{\mathrm{tw}}(\generalizedPos)$,
and $E_{\mathrm{in}}(\generalizedPos)$, $E_{\mathrm{st}}(\generalizedPos)$, $E_{\mathrm{be}}(\generalizedPos)$, and $E_{\mathrm{tw}}(\generalizedPos)$ represent inertia, stretching, bending, and twisting objectives, respectively. We define these objectives in the following sections and provide their gradient in the supplementary material.


\subsection{Inertia}
\label{sec:DER_inertia}
We define the inertia objective as follows:
\begin{align}
E_{\mathrm{in}}(\generalizedPos) =
\frac{1}{2\dt^2}\norm{\generalizedPos - \generalizedPos^*}^2_{\generalizedMass},
\end{align}
where $\generalizedPos^* = 
\generalizedPos^t + 
\dt \generalizedVel^t + 
\dt^2 \generalizedMass^{-1} \generalizedForceExt$, $\generalizedPos^t$ and $\generalizedVel^t$ denote the generalized positions and velocities at time $t$, respectively, $\dt$ the time step size, $\generalizedMass$ a diagonal generalized mass matrix \cite{Bergou:2010:DVT:1778765.1778853,jawed2018primer}, and $\generalizedForceExt$ generalized external forces. While we set gravity as the only external force, one could specify spatially varying external loads and winds (even on edge angles) \cite{Takahashi2025rest}.


\subsection{Stretching}
\label{sec:DER_stretch}
We define the stretching objective \cite{Bergou:2010:DVT:1778765.1778853} as $E_{\mathrm{st}}(\generalizedPos) = 
\sum_{i = 1}^{N-2} E_{\mathrm{st}, i}(\vertexPos_{i}, \vertexPos_{i+1})$, where $E_{\mathrm{st}, i}(\vertexPos_{i}, \vertexPos_{i+1})$ is given by
\begin{align}
E_{\mathrm{st}, i}(\vertexPos_{i}, \vertexPos_{i+1}) = 
\frac{1}{2}
\left(\frac{\stiffnessScale \stretchCoef_i \pi \radius^2}{\restLength_i}\right)
(\length_i - \restLength_i)^2,
\end{align}
with $\stiffnessScale$ denoting a global constant scalar (defined in Sec. \ref{sec:optimization_variable}) that scales stiffness parameters, $\stretchCoef_i$ as the stiffness parameter for stretching on edge $i$ (we use different stiffness parameters for stretching, bending, and twisting to enable finer control \cite{Huang2023hair,Takahashi2025rest}, unlike \cite{Bergou:2010:DVT:1778765.1778853} which uses only Young's and shear modulus), $\length_i = \norm{\vertexPos_{i+1} - \vertexPos_{i}}_2$ representing the length of edge $i$, and $\restLength_i$ as its rest length.


\subsection{Bending}
\label{sec:DER_bend}
We define the bending objective \cite{Bergou2008,Fei2019} as $E_{\mathrm{be}}(\generalizedPos) = 
\sum_{i = 1}^{N-2} E_{\mathrm{be}, i}(\yvec_i)$, where $E_{\mathrm{be}, i}(\yvec_i)$ is given by
\begin{align}
E_{\mathrm{be}, i}(\yvec_i) = 
\frac{1}{2}
\left(\frac{\stiffnessScale \bendCoef_i \pi \radius^4}{4 (\restLength_{i-1} + \restLength_{i})}\right)
\norm{\curvature_i - \restCurvature_i}_2^2,
\label{eq:bending_i}
\end{align}
and $\yvec_i = (\vertexPos_{i-1}^T, \theta_{i-1}, \vertexPos_{i}^T, \theta_{i}, \vertexPos_{i+1}^T)^T \in \realNumber^{11}$, $\bendCoef_i$ denotes the bending stiffness at vertex $i$, and $\curvature_i$ and $\restCurvature_i$ are the four-dimensional curvature and rest curvature, respectively \cite{Bergou2008,Fei2019}. We use the formulation with four-dimensional curvatures and rest curvatures \cite{Bergou2008,Fei2019} because a simplified bending formulation with two-dimensional curvatures and rest curvatures via averaged material frames \cite{Bergou:2010:DVT:1778765.1778853} is known to fail to evaluate strand bending correctly due to the non-unit and non-orthogonal averaged material frames \cite{Gornowicz2015,Panetta2019}. We provide more details on these bending formulations in the supplementary material.


\subsection{Twisting}
\label{sec:DER_twist}
We define the twisting objective \cite{Bergou:2010:DVT:1778765.1778853,jawed2018primer} as $E_{\mathrm{tw}}(\generalizedPos) = 
\sum_{i = 1}^{N-2} E_{\mathrm{tw}, i}(\yvec_i)$, where $E_{\mathrm{tw}, i}(\yvec_i)$ is given by
\begin{align}
E_{\mathrm{tw}, i}(\yvec_i) = 
\frac{1}{2}
\left(\frac{\stiffnessScale \twistCoef_i \pi \radius^4}{(\restLength_{i-1} + \restLength_{i})}\right)
(\twist_i - \restTwist_i)^2,
\label{eq:twist_i}
\end{align}
with $\twistCoef_i$ representing the twisting stiffness parameter, and $\twist_i$ and $\restTwist_i$ the twist (including the reference twist \cite{Bergou:2010:DVT:1778765.1778853,Kaldor2010,jawed2018primer}) and rest twist, respectively. 


\section{Parameter Optimization}
\label{sec:parameter_optimization}
Our goal is, at the initialization stage, to find a set of optimal rest shape and material stiffness parameters which achieve static equilibrium of the strands. Letting $\parameter$ be the optimization variable (to be detailed in Sec. \ref{sec:optimization_variable}), our parameter optimization task is formulated as a constrained minimization problem:
\begin{align}
\parameter = \argmin_{\parameterMin \leq \parameter \leq \parameterMax,
\constraint(\parameter) = 0} 
R(\parameter),
\
R(\parameter) = 
\frac{1}{2}\norm{\parameter - \parameter_{0}}_{\Wmat}^2,
\label{eq:parameter_optimization}
\end{align}
where $\parameterMin$ and $\parameterMax$ denote the box constraints (lower and upper bounds for $\parameter$, respectively) to ensure compliance with physical laws and limit significant parameter changes \cite{Takahashi2025rest}, and $\constraint(\parameter)$ (defined in Sec. \ref{sec:zero_force_constraints}) is a nonlinear hard constraint to enforce zero net forces. We define $\parameter_0$ as the initial value for $\parameter$ before optimization, and $R(\parameter)$ acts as a regularizer to minimize the deviation of $\parameter$ from $\parameter_0$ with a diagonal weight matrix $\Wmat$. We aim to solve \eqref{eq:parameter_optimization} using ALM and a Newton-type optimizer (see Algorithm \ref{alg:parameter_optimization} in Sec. \ref{sec:algorithm} for the procedure).


\subsection{Optimization Variable}
\label{sec:optimization_variable}
We simultaneously optimize parameters of the rest shape ($\restLength, \restCurvature, \restTwist$) and material stiffness ($\stretchCoef, \bendCoef, \twistCoef$) to achieve static equilibrium of an elastic strand. To improve robustness and efficiency, we use the following approaches to define the optimization variable $\parameter$.


\textbf{Variable Scaling.}
While one typically specifies stiffness coefficients directly as $\stretchCoefFinal (=\stiffnessScale \stretchCoef)$, $\bendCoefFinal(=\stiffnessScale \bendCoef)$, and $\twistCoefFinal(=\stiffnessScale \twistCoef)$, we instead optimize $\stretchCoef$, $\bendCoef$, and $\twistCoef$. This approach addresses the significant scale differences between the stiffness coefficients ($\stretchCoefFinal$, $\bendCoefFinal$, $\twistCoefFinal$ on the order of $10^9$ \cite{Bergou:2010:DVT:1778765.1778853,jawed2018primer}) and rest shape parameters ($\restLength, \restCurvature, \restTwist$ up to 10), enabling stable convergence of numerical optimizers and mitigating numerical issues. To ensure similar value scales, we initialize the stiffness scaling constant $\stiffnessScale$ as $\stiffnessScale = \frac{1}{3(N-2)}\sum_{i=1}^{N-2} (\stretchCoefFinali + \bendCoefFinali + \twistCoefFinali)$, and subsequently set $\stretchCoef = \stretchCoefFinal / \stiffnessScale$, $\bendCoef = \bendCoefFinal / \stiffnessScale$, and $\twistCoef = \twistCoefFinal / \stiffnessScale$. While equivalent formulations could be derived through proper scaling of \eqref{eq:parameter_optimization} and \eqref{eq:augmented_lagrangian}, our change of variables ensures consistent scales in the Jacobians (see the supplementary material), thus avoiding catastrophic cancellation and enhancing numerical stability. In practice, we found that parameter optimization without this change of variables fails to achieve static equilibrium.


\textbf{Variable Interleaving.}
Exploiting the one-dimensional structure and locally defined DER objectives of strands, we interleave the rest shape and stiffness parameters to ensure banded inner linear systems (which can be solved with fewer fill-ins via Cholesky-based direct solvers) within Newton-type optimizers, similar to the approach used in forward simulations \cite{Bergou:2010:DVT:1778765.1778853}. Note that we exclude $\restLength_0$ on the fixed edge from the parameter optimization \cite{Takahashi2025rest}.


\textbf{Reduced Rest Curvatures.}
Furthermore, we optimize only two rest curvature DOFs per vertex, $\restCurvature_{i, 0}$ and $\restCurvature_{i, 1}$, while excluding $\restCurvature_{i, 2}$ and $\restCurvature_{i, 3}$, in contrast to previous work \cite{Takahashi2025rest}. This choice stems from the association of $\restCurvature_{i, 0}$ and $\restCurvature_{i, 2}$ with the second material frame, while $\restCurvature_{i, 1}$ and $\restCurvature_{i, 3}$ relate to the first \cite{Bergou2008}. Their overlapping force spaces lead to similar final values for $\restCurvature_{i, 0}$ and $\restCurvature_{i, 2}$ without expanding the force spaces necessary to achieve static equilibrium, while including $\restCurvature_{i, 2}$ and $\restCurvature_{i, 3}$ would increase the optimization cost due to the additional DOFs. In practice, we synchronize these variables by setting $\restCurvature_{i, 2} = \restCurvature_{i, 0}$ and $\restCurvature_{i, 3} = \restCurvature_{i, 1}$ whenever $\restCurvature_{i, 0}$ and $\restCurvature_{i, 1}$ are updated. This approach effectively reduces the dimensionality of the rest curvature from 4D to 2D; however, we still use 4D rest curvatures in \eqref{eq:bending_i} to ensure the same dimensionality for curvature $\curvature$ and rest curvature $\restCurvature$.

Based on these choices, we define the optimization variable for the parameters as $\parameter = (
\restCurvature_{1, 0}, \restCurvature_{1, 1}, {\bendCoef}_{1},
\restTwist_1, {\twistCoef}_{1},
\restLength_1, {\stretchCoef}_{1},
\ldots,
\\
\restCurvature_{N-2, 0}, \restCurvature_{N-2, 1}, {\bendCoef}_{N-2},
\restTwist_{N-2}, {\twistCoef}_{N-2},
\restLength_{N-2}, {\stretchCoef}_{N-2}
)^T \in \realNumber^{7N-14}$. The subset corresponding to the rest shape parameters consists of $4N-8$ DOFs, which matches the number of active DOFs for the generalized positions $\generalizedPos$. Any set of rest shape parameter values can exactly be converted into any corresponding set of generalized positions (and vice versa), and thus either representation could (potentially) be optimized for. We choose to use the rest shape parameters in the optimization because this approach offers more precise control over the magnitude of forces via the box constraints, and makes the Jacobians of the generalized forces with respect to the rest shape parameters much simpler \cite{Takahashi2025rest}.


\subsection{Augmented Lagrangian Method}
To handle the nonlinear constraints $\constraint(\parameter) = 0$ in \eqref{eq:parameter_optimization}, we use ALM \cite{NoceWrig06,Skouras2012design} and define an augmented Lagrangian objective with a Lagrange multiplier $\multiplier$ and a penalty parameter $\penaltyCoef \ (>0)$:
\begin{align}
L(\parameter, \multiplier) = 
R(\parameter) -
\multiplier^T \constraint(\parameter) +
\frac{\penaltyCoef}{2}\norm{\constraint(\parameter)}_2^2.
\label{eq:augmented_lagrangian}
\end{align}
We then reformulate the parameter optimization \eqref{eq:parameter_optimization} as
\begin{align}
\parameter, \multiplier = \argmin_{\parameterMin \leq \parameter \leq \parameterMax}\argmax_{\multiplier}
L(\parameter, \multiplier),
\label{eq:parameter_optimization_alm}
\end{align}
which we optimize by iteratively solving primal and dual subproblems \cite{NoceWrig06}. The primal minimization subproblem, with fixed $\multiplier^{\iterIndex}$ for iteration index $\iterIndex$ (initialized with $\multiplier^{0} = 0$), is defined as
\begin{align}
\parameter^{\iterIndex + 1} = \argmin_{\parameterMin \leq \parameter \leq \parameterMax}
L(\parameter, \multiplier^{\iterIndex}),
\label{eq:primal_optimization}
\end{align}
while the dual maximization subproblem can be solved via a simple vector update:
\begin{align}
\multiplier^{\iterIndex + 1} = \multiplier^{\iterIndex} - 
\penaltyCoef \constraint(\parameter^{\iterIndex+1}).
\label{eq:dual_update}
\end{align}
Thus, efficiently solving the primal subproblem \eqref{eq:primal_optimization} is crucial for fast parameter optimization.


\subsection{Zero Net Force Constraints}
\label{sec:zero_force_constraints}
Given the total force due to the DER objectives, represented as $\generalizedForce(\parameter) = -\nabla E(\generalizedPos)$, it seems natural to define $\constraint(\parameter) = \generalizedForce(\parameter)$ to enforce zero net forces. However, in this case, the term $\frac{\penaltyCoef}{2}\norm{\constraint(\parameter)}_2^2$ in \eqref{eq:augmented_lagrangian} becomes equivalent to a quadratic penalty on the total force, which is numerically ill-conditioned and fails to correctly account for the force contributions within the system \cite{Takahashi2025rest}. To avoid these issues, we penalize the generalized forces in the kinetic energy metric \cite{Takahashi2025rest} and define $\constraint(\parameter)$ as
\begin{align}
\constraint(\parameter) = \generalizedMass^{-\frac{1}{2}} \generalizedForce(\parameter).
\label{eq:constraint}
\end{align}
Consequently, $L(\parameter, \multiplier)$ in \eqref{eq:augmented_lagrangian} can be rewritten as
\begin{align}
L(\parameter, \multiplier) = 
R(\parameter) -
\multiplier^T \generalizedMass^{-\frac{1}{2}}\generalizedForce(\parameter) +
\frac{\penaltyCoef}{2}
\norm{\generalizedForce(\parameter)}^2_{\generalizedMass^{-1}}.
\label{eq:augmented_lagrangian2}
\end{align}
This objective $\eqref{eq:augmented_lagrangian2}$ is numerically equivalent to the formulation proposed by Takahashi and Batty \cite{Takahashi2025rest} when considering only the rest shape parameters (excluding stiffness parameters) and setting $\multiplier = 0$. Thus, their formulation, which treats $\constraint(\parameter)$ as a soft constraint, can be regarded as a subset of ours. In the following, we assume $\dt = 1$ to simplify the formulations without affecting the optimization results \cite{Takahashi2025rest}.


\vspace{-1mm}
\subsection{Box Constraints}
We impose box constraints on the optimization parameters to prevent stability problems (in forward simulation) which can be caused by too significant changes in the parameter values. Specifically, we define
$\restLength_{\mathrm{min}} \leq \restLength \leq \restLength_{\mathrm{max}}$ (except for $\restLength_0$), $\restCurvature_{\mathrm{min}} \leq \restCurvature \leq \restCurvature_{\mathrm{max}}$, 
$\restTwist_{\mathrm{min}} \leq \restTwist \leq \restTwist_{\mathrm{max}}$,
$\stretchCoef_{\mathrm{min}} \leq \stretchCoef \leq \stretchCoef_{\mathrm{max}}$, 
$\bendCoef_{\mathrm{min}} \leq \bendCoef \leq \bendCoef_{\mathrm{max}}$, and
$\twistCoef_{\mathrm{min}} \leq \twistCoef \leq \twistCoef_{\mathrm{max}}$. Unlike the penalty-based approach \cite{Takahashi2025rest}, which requires safety margins to mitigate the risk of violating physical laws (e.g., negative rest lengths and material parameters), we handle box constraints precisely using MPRGP \cite{Dostal2005}. Consequently, we set lower and upper bounds for $\restLength$, $\stretchCoef$, $\bendCoef$, and $\twistCoef$ as
\begin{align}
\restLength_{\mathrm{min}} &= \epsilon, \quad
\restLength_{\mathrm{max}} = \infty, \quad
\stretchCoef_{\mathrm{min}} = \epsilon, \quad
\stretchCoef_{\mathrm{max}} = \infty,
\\
\bendCoef_{\mathrm{min}} &= \epsilon, \quad
\bendCoef_{\mathrm{max}} = \infty, \quad
\twistCoef_{\mathrm{min}} = \epsilon, \quad
\twistCoef_{\mathrm{max}} = \infty,
\end{align}
where $\epsilon$ denotes a small positive value (we set $\epsilon = 10^{-10}$). While we set the upper bounds of stiffness parameters to $\infty$ to ensure that static equilibrium is achievable \cite{Derouet-Jourdan2010,Hsu2022sag}, finite values can be used if one prefers to avoid excessively stiff materials that may compromise static equilibrium.

We set the lower/upper bounds for $\restCurvature$ and $\restTwist$ as
\begin{align}
\restCurvature_{\mathrm{min}} &= \restCurvature_{0} - \curvRange \evec_{\mathrm{be}},
\quad
\restCurvature_{\mathrm{max}} = \restCurvature_{0} + \curvRange \evec_{\mathrm{be}},
\\
\restTwist_{\mathrm{min}} &= \restTwist_{0} - \eta \evec_{\mathrm{tw}},
\quad
\restTwist_{\mathrm{max}} = \restTwist_{0} + \eta \evec_{\mathrm{tw}},
\end{align}
where $\restCurvature_0$ and $\restTwist_0$ denote the initial rest curvature and rest twist (before parameter optimization), respectively, $\curvRange$ and $\eta$ the allowed change for rest curvature and rest twist, respectively, and $\evec_{\mathrm{be}}$ and $\evec_{\mathrm{tw}}$ vectors of all ones with the same dimensions as $\restCurvature$ and $\restTwist$, respectively. The stable range of $\restCurvature$ (i.e., which does not cause stability problems in forward simulations) and thus $\curvRange$ can vary depending on simulation settings (e.g., strand geometry and material stiffness) \cite{Takahashi2025rest}; we use $\curvRange = 1$ by default based on our experiments. Additionally, since changes in rest twist are typically much smaller than those in the rest curvature, we set $\eta = \curvRange / 4$ to reduce the number of tunable parameters.


\subsection{Newton-based Box-Constrained Optimization}
We solve the primal, box-constrained nonlinear minimization subproblem \eqref{eq:primal_optimization} using a Newton-type optimizer. While one could approximate the box constraints with a penalty objective to ensure fully unconstrained inner linear systems \cite{Takahashi2025rest}, penalty methods often require tedious parameter tuning to achieve acceptable results \cite{NoceWrig06}. Even with carefully tuned parameters, these methods may lead to compromises that violate box constraints. Therefore, instead of relying on penalty methods, we incorporate box constraints directly when computing the Newton directions, which is equivalent to solving box-constrained quadratic programs (BCQPs) \cite{Takahashi2021}.

The Newton direction $\Delta \parameter$ at $\iterIndex + 1$ iteration can be computed by solving the following BCQP:
\begin{align}
\Delta \parameter^{\iterIndex + 1} &= 
\argmin_{
\Delta \parameter_{\mathrm{min}}^{\iterIndex} \leq 
\Delta \parameter \leq 
\Delta \parameter_{\mathrm{max}}^{\iterIndex}
}
F(\Delta \parameter),
\label{eq:bcqp}
\\
F(\Delta \parameter) &=
\frac{1}{2} \Delta \parameter^T 
\nabla^2 
L (\parameter^{\iterIndex})
\Delta \parameter 
+
\Delta \parameter^T 
\nabla 
L (\parameter^{\iterIndex}, \multiplier^{\iterIndex}),
\end{align}
where $\Delta \parameter_{\mathrm{min}}^{\iterIndex}$ and $\Delta \parameter_{\mathrm{max}}^{\iterIndex}$ denote lower and upper bounds for $\Delta \parameter$, respectively. We omit $\multiplier^{\iterIndex}$ in the argument of $\nabla^2 L(\parameter)$ as it is independent of $\multiplier^{\iterIndex}$ \eqref{eq:hessian}. Assuming the ideal step length of one \cite{NoceWrig06}, we have $\parameterMin \leq \parameter^{\iterIndex+1} = \parameter^{\iterIndex} + \Delta \parameter^{\iterIndex + 1} \leq \parameterMax$, giving $\Delta \parameter_{\mathrm{min}}^{\iterIndex} = \parameterMin - \parameter^{\iterIndex}$ and $\Delta \parameter_{\mathrm{max}}^{\iterIndex} = \parameterMax - \parameter^{\iterIndex}$ \cite{Takahashi2021}.

Given the augmented Lagrangian objective \eqref{eq:augmented_lagrangian2}, we can compute its gradient as
\begin{align}
\nabla L(\parameter, \multiplier^{\iterIndex}) =
\nabla R(\parameter) -
\Jacobian^T \generalizedMass^{-\frac{1}{2}} \multiplier^{\iterIndex} +
\penaltyCoef \Jacobian^T \generalizedMass^{-1} \generalizedForce(\parameter),
\label{eq:gradient}
\end{align}
where $\nabla R(\parameter) = \Wmat(\parameter - \parameter_{0})$, and we denote the Jacobian of $\generalizedForce(\parameter)$ with respect to $\parameter$ by $\Jacobian \left(=\frac{\partial \generalizedForce(\parameter)}{\partial \parameter}\right) \in \realNumber^{(4N-8) \times (7N-14)}$  (details are given in the supplementary material). Note that we treat $\generalizedMass$ (which is computed with the initial rest length $\restLength$ \cite{Bergou:2010:DVT:1778765.1778853,jawed2018primer}) as constant, as $\generalizedMass$ is introduced and intended to properly scale the generalized forces (see \eqref{eq:constraint}) \cite{Takahashi2025rest}.

Since the generalized forces are linear with respect to the rest curvature, rest twist, and stiffness parameters, the second order derivatives with respect to these parameters are zero. While the second order derivatives with respect to $\restLength$ are non-zero, these components would be projected to zero to ensure SPD inner systems and valid Newton descent directions \cite{NoceWrig06}. In addition, the change in $\restLength$ during the parameter optimization is typically small for less extensible strands, i.e., $\restLength$ can be close to constant. Thus, using the exact Hessian with SPD projections does not have discernible benefits while increasing the implementation complexity and computational cost due to the involvement of third order tensors, as demonstrated and discussed in \cite{Takahashi2025rest}. Therefore, we ignore the second order derivatives with respect to $\restLength$ and approximate the Hessian in a Gauss-Newton style (excluding $\multiplier$ from arguments) by
\begin{align}
\nabla^2 L(\parameter) \approx
\Wmat + \penaltyCoef 
\Jacobian^T \generalizedMass^{-1} \Jacobian.
\label{eq:hessian}
\end{align}
As this approximated Hessian is guaranteed to be SPD, we can solve the convex BCQP \eqref{eq:bcqp} using MPRGP \cite{Dostal2005}, accelerated with our active-set Cholesky preconditioner (see Sec. \ref{sec:bcqp_solver}).


\subsection{Algorithm and Implementation}
\label{sec:algorithm}
Algorithm \ref{alg:parameter_optimization} outlines our parameter optimization method. To accelerate convergence and reduce the risk of converging to undesirable local minima, we use warm starting. We set $\penaltyCoef = 10^6$, and use $10^{3}$ and $10^{0}$ for the stiffness and rest shape parts of diagonals in $\Wmat$, respectively, so that we prioritize achieving (in order) zero net forces, small changes in the stiffness parameters, and then in rest shape parameters. We perform a single Newton iteration to solve our primal BCQP subproblem \eqref{eq:primal_optimization} as it serves as an inner problem within \eqref{eq:parameter_optimization_alm}. To guarantee a decrease in the objective, we implement a backtracking line search \cite{NoceWrig06}. The iterations are terminated when $\norm{\Delta \parameter}_2$ falls below a threshold $\epsilon_p \ (= 10^{-8})$ or iteration count exceeds $\iterIndex_{\mathrm{max}} \ (= 100)$. In addition, we can also terminate the solver iterations to prevent infinite loops if we cannot make any progress with the backtracking line search (although our method did not experience this case in our examples).

\begin{algorithm}
\caption{Parameter Optimization}
\label{alg:parameter_optimization}
\begin{algorithmic}[1]
\STATE Initialize generalized positions $\generalizedPos$, radius $\radius$, stiffness coefficients $\stretchCoefFinal$, $\bendCoefFinal$, $\twistCoefFinal$, length $\length$, rest length $\restLength$, generalized mass matrix $\generalizedMass$, unit tangent vector $\tangent$, reference and material frames, curvature $\curvature$, rest curvature $\restCurvature$, twist $\twist$, rest twist $\restTwist$, stiffness scaling $\stiffnessScale$, and material stiffness parameters $\stretchCoef$, $\bendCoef$, $\twistCoef$, penalty parameter $\penaltyCoef$, weight matrix $\Wmat$, Lagrange multiplier $\multiplier^{\iterIndex} = 0$, iteration index $\iterIndex = 0$
\STATE Set $\parameter_0 = \parameter$ and compute $\parameterMin$ and $\parameterMax$
\STATE \textbf{do}
\STATE \quad Compute box constraints $\Delta \parameterMin^{\iterIndex}$ and $\Delta \parameterMax^{\iterIndex}$
\STATE \quad Compute $\nabla L(\parameter^{\iterIndex}, \multiplier^{\iterIndex})$, $\nabla^2 L(\parameter^{\iterIndex})$ with \eqref{eq:gradient} and \eqref{eq:hessian}
\STATE \quad Solve $\nabla^2 L(\parameter^{\iterIndex}) \Delta \parameter^{\iterIndex + 1} = -\nabla L(\parameter^{\iterIndex}, \multiplier^{\iterIndex})$ with $\Delta \parameterMin^{\iterIndex}$ and $\Delta \parameterMax^{\iterIndex}$
\STATE \quad Update to $\parameter^{\iterIndex+1}$ using backtracking line search with \eqref{eq:augmented_lagrangian2} and synchronization for $\restCurvature$
\STATE \quad Update Lagrange multipliers with \eqref{eq:dual_update}
\STATE \quad $\iterIndex = \iterIndex + 1$
\STATE \textbf{while} $\epsilon_p < \norm{\Delta \parameter^{k+1}}_2$ and $\iterIndex < \iterIndex_{\mathrm{max}}$
\end{algorithmic}
\end{algorithm}



\section{Box-Constrained Quadratic Program Solver}
\label{sec:bcqp_solver}
Box constraints are a specific type of inequality constraint that can be handled efficiently through active-set methods, without relying on barriers or penalties \cite{NoceWrig06}. Since our BCQPs arise as primal subproblems \eqref{eq:primal_optimization} within the constrained nonlinear minimization \eqref{eq:parameter_optimization_alm}, we opt to solve them with a prescribed level of accuracy suitable for inexact Newton-like methods. This approach enables us to conserve computational resources by avoiding unnecessary precision \cite{NoceWrig06}. Thus, instead of employing direct solvers with active sets (e.g., Lemke's method or Dantzig's simplex algorithm), we prefer iterative methods that permit early termination. Given the stiff yet SPD inner systems, we employ MPRGP \cite{Dostal2005} and propose an effective new preconditioner based on full Cholesky factorization and triangular solves with active sets, utilizing the banded system structures. We refer to this scheme as active-set Cholesky (ASC) preconditioning. Notably, since the Cholesky-based approach precisely solves a linear system, Cholesky-preconditioned MPRGP (or CG) finds a solution in just one iteration when no box constraints are activated.


\subsection{Active-Set Cholesky Preconditioning}
\label{sec:active_set_cholesky}
Consider minimizing a quadratic objective $\frac{1}{2}\xvec^T \Amat \xvec - \xvec^T \bvec$ and corresponding linear system $\Amat \xvec = \bvec$. The associated linear preconditioning system can be expressed as $\Amat \zvec = \rvec$, where $\zvec$ and $\rvec$ are vectors corresponding to $\xvec$ and $\bvec$, respectively. Adopting an active set $\avec$ that indicates whether $\xvec$ is limited or not (i.e., with an index $i$, $\avec_i = 1$ or $\avec_i = -1$ if $\xvec_i$ is limited by its lower or upper bound, respectively, and $\avec_i = 0$ otherwise) \cite{Takahashi2023}, we can define a diagonal selection matrix $\Smat$ with $\Smat_{ii} = 1 - |\avec_i|$, which can also be interpreted as a filtering matrix. Given $\zvec_u$ and $\zvec_c$ that denote unconstrained and constrained parts of $\zvec$, respectively, we need to solve the preconditioning system for $\zvec_u$ while enforcing $\zvec_c = 0$ \cite{Takahashi2023}. The preconditioning system can be rewritten (incorporating the necessary constraints \cite{Tamstorf:2015:SAM:2816795.2818081}) as 
$(\Smat \Amat \Smat^T + \Imat - \Smat) \zvec = \Smat \rvec$, or given equivalently (by splitting it for $\zvec_u$ and $\zvec_c$) as 
$\Smat_{u}\Amat \Smat_{u}^T \zvec_{u} = \Smat_u \rvec$ and $\zvec_{c} = 0$,
where $\Smat_u$ denotes a matrix consisting of $\Smat$'s rows associated with the unconstrained variables.


\subsubsection{Problems with Naive Application of Cholesky Solver}
While one approach to solving these preconditioning systems is to reassemble $(\Smat \Amat \Smat^T + \Imat - \Smat)$ or $\Smat_{u}\Amat \Smat_{u}^T$ and factorize it using Cholesky decomposition, this procedure is costly because active sets can change in each MPRGP iteration \cite{Takahashi2023}, and thus it is ideal to perform Cholesky decomposition only once and reuse its resulting factor.

We consider Cholesky factorization $\Amat = \Lmat \Lmat^T$ (where $\Lmat$ is a lower triangular matrix). While its naive substitution to $(\Smat \Amat \Smat^T + \Imat - \Smat) \zvec = \Smat \rvec$ gives $(\Smat \Lmat \Lmat^T \Smat^T + \Imat - \Smat) \zvec = \Smat \rvec$, triangular solves are inapplicable to this form. Similarly, while $\Smat_{u}\Amat \Smat_{u}^T \zvec_{u} = \Smat_u \rvec$ can be transformed into $\Smat_{u} \Lmat \Lmat^T \Smat_{u}^T \zvec_{u} = \Smat_u \rvec$, we cannot perform triangular solves with $\Smat_u \Lmat \yvec = \Smat_u \rvec$ and $\Lmat^T \Smat_u^T \zvec_u = \yvec$ since $\Smat_u \Lmat$ is generally not square. Another attempt is to solve $\Smat \Amat \Smat^T \zvec = \Smat \rvec$ (and thus $\Smat \Lmat \Lmat^T \Smat^T \zvec = \Smat \rvec$) while enforcing $\zvec_c = 0$ (e.g., by setting $\zvec_c$ to zero during the triangular solves). With this form, the forward substitution $\Smat \Lmat \yvec = \Smat \rvec$ can be written in the elementwise notation as $
\yvec_i = \frac{\Smat_{ii} \rvec_i - 
\Smat_{ii} \sum_{j < i} \Lmat_{ij} \yvec_j}{\Smat_{ii} \Lmat_{ii}}$,
and back substitution $\Lmat^T \Smat^T \zvec = \yvec$ as
$
\zvec_i = \frac{\yvec_i - 
\sum_{j > i} \Smat_{jj} \Lmat_{ji} \zvec_j}{\Smat_{ii} \Lmat_{ii}}$.
However, this approach still has two problems. First, $\Smat_{ii} = 0$ for constrained variables, rendering these operations infeasible. Second, the forward and back substitution are asymmetric, violating the symmetry requirement for preconditioning in symmetric Krylov iterative solvers, such as CG and MPRGP \cite{Saad2003}.


\subsubsection{Our Preconditioning}
Considering the equivalence between forward/back substitution and forward/backward GS applied to lower/upper triangular matrices, we can interpret triangular solves as SGS. Thus, instead of directly filtering the system matrix $\Amat$ with $\Smat$, we perform the filtering in SGS to solve a system equivalent to the filtered preconditioning system. Given the elementwise form of forward GS, $\yvec_i = (\rvec_i - \sum_{j<i} \Lmat_{ij} \yvec_j)/\Lmat_{ii}$, we can filter this operation with $\Smat$ while ensuring symmetry by
\begin{align}
\yvec_i = 
(\Smat_{ii} \rvec_i - 
\Smat_{ii} \sum_{j < i} \Smat_{jj} \Lmat_{ij} \yvec_j) / \Lmat_{ii}.
\label{eq:forward_filtered}
\end{align}
Similarly, given $\zvec_i = (\yvec_i - \sum_{j>i} \Lmat_{ji} \zvec_j)/\Lmat_{ii}$ for backward GS, we add filtering to obtain
\begin{align}
\zvec_i = (\Smat_{ii} \yvec_i - 
\Smat_{ii} \sum_{j > i} \Smat_{jj} \Lmat_{ji} \zvec_j)/\Lmat_{ii}.
\label{eq:backward_filtered}
\end{align}
Note that it is unnecessary to filter the denominator $\Lmat_{ii}$ in \eqref{eq:forward_filtered} and \eqref{eq:backward_filtered} because $\yvec$ and $\zvec$ are properly filtered in the numerator. Our filtered SGS preserves the symmetry of operations required for preconditioning of CG/MPRGP \cite{Saad2003}, and can be rewritten in the block form as $(\tDmat + \Smat \tLmat \Smat^T)\yvec = \Smat \rvec$ and $(\tDmat + \Smat^T \tLmat^T \Smat)\zvec = \Smat \yvec$, respectively, where $\Lmat = \tDmat + \tLmat$, and $\tDmat$ and $\tLmat$ denote the diagonal and strictly lower parts of $\Lmat$, respectively. Note that while \eqref{eq:forward_filtered} and \eqref{eq:backward_filtered} are equivalent to SGS (except for filtering), our $\Lmat$ arises from Cholesky factorization, in contrast to the strictly lower triangular matrix $\hLmat$ from the traditional SGS (where $\Amat = \hat{\Dmat} + \hLmat + \hLmat^T$, and $\hat{\Dmat}$ is the diagonal part of $\Amat$). Additionally, while it is typical to initialize $\yvec = 0$ and $\zvec = 0$ in GS-type preconditioning \cite{Saad2003}, our scheme does not require such initialization because \eqref{eq:forward_filtered} and \eqref{eq:backward_filtered} directly overwrite $\yvec$ and $\zvec$.

In practice, we prefer sqrt-free Cholesky factorization $\Amat = \Lmat \Dmat \Lmat^T$, where $\Lmat$ is unit lower triangular (for clarity, we redefine $\Lmat$ here), and $\Dmat$ is  diagonal. This approach can detect negative pivots (to clamp them for robustness) and avoid sqrt operations for efficiency \cite{Huang2023hair}. The triangular solve proceeds as $\Lmat \wvec = \rvec$, $\Dmat \yvec = \wvec$, and $\Lmat^T \zvec = \yvec$. By merging the forward substitution and diagonal scaling (which preserves symmetry) and skipping unnecessary computations (e.g., multiplications with $\Smat_{ii}$ and $\Lmat_{ii} (= 1)$, and scanning rows in $\Lmat$), we have forward and backward operations for $\yvec$ and $\zvec$, respectively, as
\begin{align}
\yvec_i &= 
\begin{cases} 
(\rvec_i - \sum_{j < i} \Smat_{jj} \Lmat_{ij} \yvec_j) / \Dmat_{ii} & \Smat_{ii} \neq 0, \\
  0 & \mathrm{otherwise},
\end{cases}
\\
\zvec_i &=
\begin{cases} 
\yvec_i - \sum_{j > i} \Smat_{jj} \Lmat_{ji} \zvec_j & \Smat_{ii} \neq 0, \\
  0 & \mathrm{otherwise}.
\end{cases}
\end{align}


\subsection{Active-Set Incomplete Cholesky Preconditioning}
\label{sec:incompleteCholesky}
Our preconditioning strategy utilizing active sets can also be applied to IC preconditioning. Notably, when the system is banded and fully populated within the band, Cholesky and IC factorization are equivalent, making their preconditioning (even with active sets) identical.

In our parameter optimization, while the system is banded, it is not entirely filled because, e.g., optimization variables for rest curvatures and stretching stiffness are not coupled (i.e., the Hessian \eqref{eq:hessian} lacks off-diagonal elements relating them). Cholesky decomposition is guaranteed to succeed for SPD systems (except for the case where pivot elements become negative due to numerical error \cite{Herholz2018}), whereas IC factorization may fail unless the system matrix possesses certain properties, such as being an M-matrix \cite{Chen2021multiscale}. This necessitates reordering the system (i.e., pivoting) potentially breaking the banded structures, adding positive diagonals \cite{Chen2021multiscale}, or reverting to GS \cite{Bridson2007course}, ultimately reducing preconditioning effectiveness. Given that approximately 97\% of the band is filled, it is acceptable to introduce a small number of fill-ins, considering the IC issues above. We therefore prefer full Cholesky factorization.


\section{Results and Discussions}
We implemented our method in C++20 with double-precision floating-point for scalar values and parallelized forward simulation and parameter optimization with OpenMP, processing each strand concurrently. We executed all the examples on a desktop machine with an Intel Core i7-9700 (8 cores) with 16GB RAM. For forward simulation, we use the exact Hessian with SPD projection for stretching objective and Gauss-Newton Hessian approximation for bending and twisting objectives \cite{Shi2023curls}. We perform a single Newton iteration per simulation step \cite{Huang2023hair}, executing four simulation steps per frame, except for Figures \ref{fig:teaser} and \ref{fig:hairs}, which used 12 simulation steps per frame. We use 60 frames per second. For MPRGP, we use a termination absolute or relative residual of $10^{-10}$. Unless specified otherwise, we use $\stretchCoefFinal = 10^{9}, \bendCoefFinal = 10^{9}, \twistCoefFinal = 10^{9}$. Material frames are rendered at the centers of the corresponding edges in red and yellow, with lengths matching the rest lengths of the edges.


\begin{figure}
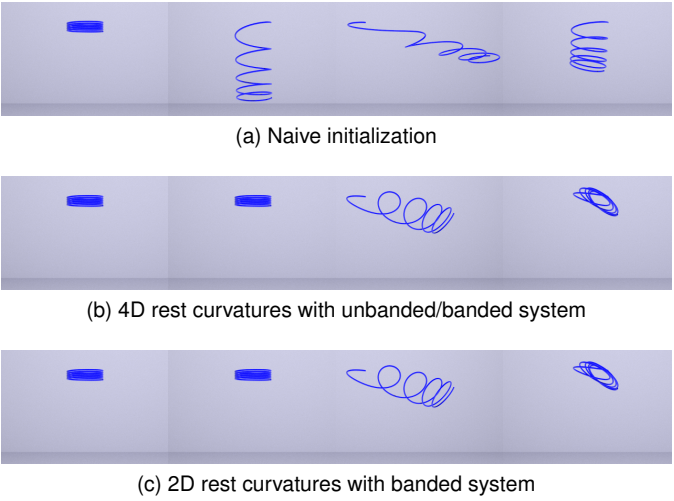
 
\centering
\subfloat[Naive initialization\label{1a}]{%
    \adjincludegraphics[trim={{0.25\width} {0.3\height} {0.25\width} {0.1\height}}, clip, width=0.25\linewidth]
    {figures/\figDir/banded_curv2_unbanded4_0001.png}%
    \adjincludegraphics[trim={{0.25\width} {0.3\height} {0.25\width} {0.1\height}}, clip, width=0.25\linewidth]
    {figures/\figDir/banded_curv2_unbanded4_0025.png}%
    \adjincludegraphics[trim={{0.5\width} {0.3\height} {0.0\width} {0.1\height}}, clip, width=0.25\linewidth]
    {figures/\figDir/banded_curv2_unbanded4_0188.png}%
    \adjincludegraphics[trim={{0.25\width} {0.3\height} {0.25\width} {0.1\height}}, clip, width=0.25\linewidth]
    {figures/\figDir/banded_curv2_unbanded4_0600.png}%
}%
\\
\subfloat[4D rest curvatures with unbanded/banded system\label{1a}]{%
    \adjincludegraphics[trim={{0.25\width} {0.3\height} {0.25\width} {0.1\height}}, clip, width=0.25\linewidth]
    {figures/\figDir/banded_curv2_banded4_0001.png}%
    \adjincludegraphics[trim={{0.25\width} {0.3\height} {0.25\width} {0.1\height}}, clip, width=0.25\linewidth]
    {figures/\figDir/banded_curv2_banded4_0025.png}%
    \adjincludegraphics[trim={{0.5\width} {0.3\height} {0.0\width} {0.1\height}}, clip, width=0.25\linewidth]
    {figures/\figDir/banded_curv2_banded4_0188.png}%
    \adjincludegraphics[trim={{0.25\width} {0.3\height} {0.25\width} {0.1\height}}, clip, width=0.25\linewidth]
    {figures/\figDir/banded_curv2_banded4_0600.png}%
}%
\\
\subfloat[2D rest curvatures with banded system\label{1a}]{%
    \adjincludegraphics[trim={{0.25\width} {0.3\height} {0.25\width} {0.1\height}}, clip, width=0.25\linewidth]
    {figures/\figDir/banded_curv2_banded2_0001.png}%
    \adjincludegraphics[trim={{0.25\width} {0.3\height} {0.25\width} {0.1\height}}, clip, width=0.25\linewidth]
    {figures/\figDir/banded_curv2_banded2_0025.png}%
    \adjincludegraphics[trim={{0.5\width} {0.3\height} {0.0\width} {0.1\height}}, clip, width=0.25\linewidth]
    {figures/\figDir/banded_curv2_banded2_0188.png}%
    \adjincludegraphics[trim={{0.25\width} {0.3\height} {0.25\width} {0.1\height}}, clip, width=0.25\linewidth]
    {figures/\figDir/banded_curv2_banded2_0600.png}%
}%
\caption{
Coil-like strand test. Rest shape optimization achieves static equilibrium unlike naive initialization (a). 4D rest curvatures with unbanded/banded systems generate identical results (b), while the reduced 2D rest curvatures also produce equivalent results given the redundant force spaces (c).
}
\label{fig:banded} 
\end{figure}

\subsection{Banded System with Reduced Rest Curvatures}
To assess the performance improvement achieved through banded systems and reduced rest curvatures, we experiment with a coil-like strand discretized with 2k vertices, as shown in Figure \ref{fig:banded} (for reference, we include naive initialization, which computes rest shape parameters with the generalized positions of the input strand). For a fair comparison with previous work \cite{Takahashi2025rest}, we optimize only the rest shape parameters and exclude box constraints to eliminate the influence of active sets. We compare the following:
\begin{enumerate}
\item Unbanded system with 4D rest curvatures \cite{Takahashi2025rest};
\item Banded system with 4D rest curvatures;
\item Banded system with 2D rest curvatures.
\end{enumerate}

Using the approximate minimum degree (AMD) reordering, the sequential arrangement of the rest shape parameters (rest length, rest curvatures, and rest twist) \cite{Takahashi2025rest} aligns with the banded system, yielding identical results. The AMD reordering took 2.0 ms while the system solve took 5.0 ms. As the banded system (with 4D rest curvatures) eliminates the need for the reordering, we achieve around $29\%$ performance gain.

Due to the redundant force spaces with rest curvatures, the virtually reduced 2D rest curvatures produce results comparable to those obtained with 4D rest curvatures. The computation times were 2.1 ms and 5.0 ms with 2D and 4D rest curvatures, respectively. As the DOF count is $4N-8$ and $6N-12$, with non-zero counts per row of $22.0$ and $32.0$ on average, the total number of non-zeros is approximately $88N (=4N \times 22)$ and $192N (=6N \times 32)$ for 2D and 4D rest curvatures, respectively. Thus, the performance ratio $0.42 = (2.1 / 5.0)$ is in good agreement with the ratio of the total number of non-zeros $0.46 \approx (88N / 192N)$.


\begin{figure}
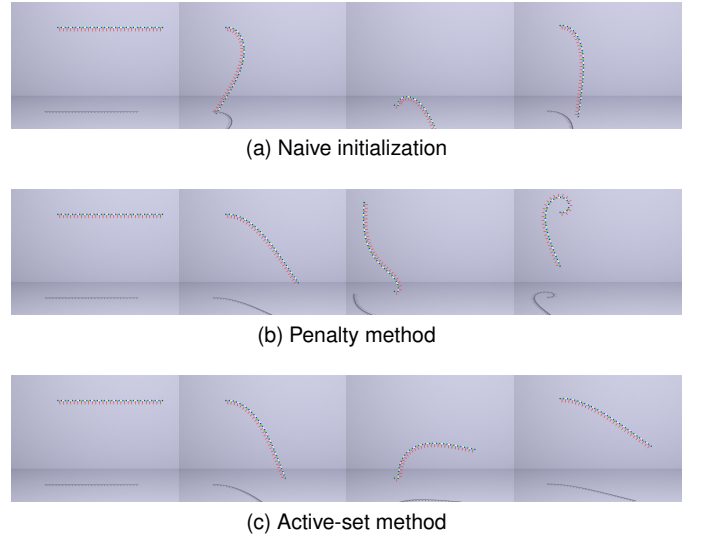
 
\centering
\subfloat[Naive initialization\label{1a}]{%
    \adjincludegraphics[trim={{0.05\width} {0.0\height} {0.2\width} {0.0\height}}, clip, width=0.25\linewidth]
    {figures/\figDir/penalty_naive_0001.png}%
    \adjincludegraphics[trim={{0.05\width} {0.0\height} {0.2\width} {0.0\height}}, clip, width=0.25\linewidth]
    {figures/\figDir/penalty_naive_0050.png}%
    \adjincludegraphics[trim={{0.05\width} {0.0\height} {0.2\width} {0.0\height}}, clip, width=0.25\linewidth]
    {figures/\figDir/penalty_naive_0225.png}%
    \adjincludegraphics[trim={{0.05\width} {0.0\height} {0.2\width} {0.0\height}}, clip, width=0.25\linewidth]
    {figures/\figDir/penalty_naive_0450.png}%
}%
\\
\subfloat[Penalty method\label{1a}]{%
    \adjincludegraphics[trim={{0.05\width} {0.0\height} {0.2\width} {0.0\height}}, clip, width=0.25\linewidth]
    {figures/\figDir/penalty_penalty_0001.png}%
    \adjincludegraphics[trim={{0.05\width} {0.0\height} {0.2\width} {0.0\height}}, clip, width=0.25\linewidth]
    {figures/\figDir/penalty_penalty_0050.png}%
    \adjincludegraphics[trim={{0.05\width} {0.0\height} {0.2\width} {0.0\height}}, clip, width=0.25\linewidth]
    {figures/\figDir/penalty_penalty_0225.png}%
    \adjincludegraphics[trim={{0.05\width} {0.0\height} {0.2\width} {0.0\height}}, clip, width=0.25\linewidth]
    {figures/\figDir/penalty_penalty_0450.png}%
}%
\\
\subfloat[Active-set method\label{1a}]{%
    \adjincludegraphics[trim={{0.05\width} {0.0\height} {0.2\width} {0.0\height}}, clip, width=0.25\linewidth]
    {figures/\figDir/penalty_activeset_0001.png}%
    \adjincludegraphics[trim={{0.05\width} {0.0\height} {0.2\width} {0.0\height}}, clip, width=0.25\linewidth]
    {figures/\figDir/penalty_activeset_0050.png}%
    \adjincludegraphics[trim={{0.05\width} {0.0\height} {0.2\width} {0.0\height}}, clip, width=0.25\linewidth]
    {figures/\figDir/penalty_activeset_0225.png}%
    \adjincludegraphics[trim={{0.05\width} {0.0\height} {0.2\width} {0.0\height}}, clip, width=0.25\linewidth]
    {figures/\figDir/penalty_activeset_0450.png}%
}%
\caption{
\mycolor{We vertically translate the root position (initially the leftmost endpoint) of a horizontal strand up and down. The leftmost column shows the initialized states for each method and the remaining columns from left to right show the results of forward simulation.} The penalty method fails to precisely enforce box constraints, causing significant rest curvature changes and allowing the strand's tail to end up on the left side after \mycolor{the motion stops} (b). By contrast, our active-set method (c) strictly enforces the box constraints, keeping the tail on the right side and better preserving the original shape than the naive initialization (a).
}
\label{fig:activeset} 
\end{figure}

\subsection{Penalty vs. Active-Set Methods}
\label{sec:penalty_active_set}
To demonstrate the efficacy of our active-set-based BCQP solver, we compare it to a penalty-based one, using a horizontal strand discretized with 30 vertices, as shown in Figure \ref{fig:activeset} (naive initialization included for reference). In this comparison, we optimize rest shape parameters only and use $\curvRange = 0.4$ to prevent significant rest curvature changes and thus to keep the tail end of the strand on the right. For the penalty method, we use a Cholesky solver \cite{Takahashi2025rest}. As we perform only one Newton iteration to solve each BCQP, if we apply ALM (with Lagrange multipliers initialized to zero) to the BCQP, it becomes equivalent to the penalty method \cite{NoceWrig06}.

The penalty method failed to strictly enforce the rest curvature changes within their corresponding box constraints. Consequently, significant rest shape changes caused the tail end of the strand to flip to the left side after the root was perturbed. By contrast, our active-set method precisely constrains curvature changes via their box constraints, keeping the tail end of the strand on the right side. While we may need to compromise static equilibrium, our method better preserves the original shape compared to naive initialization.

The penalty method uses 4 Newton iterations with a total system-solving cost of 0.31 ms (0.078 ms per solve), whereas our method uses 2 Newton iterations with 8.0 MPRGP iterations on average per solve, resulting in a total cost of 0.33 ms (0.17 ms per solve). Although MPRGP needs multiple iterations, our method performs Cholesky factorization (which is costlier than triangular solves) only once per Newton iteration, leading to a moderate cost per MPRGP solve. Furthermore, the strict enforcement of box constraints enhances the convergence of Newton iterations. Consequently, the total cost of our method is comparable to that of the penalty-based approach.


\subsection{Simultaneous Optimization of Rest Shape and Material Stiffness Parameters Under Hard Zero Net Force Constraints}
To demonstrate the necessity of optimizing both rest shape and material stiffness parameters and enforcing $\constraint(\parameter) = 0$ as a hard constraint, we compare the following schemes:
\begin{enumerate}
\item Rest shape only: rest shape only optimization \cite{Takahashi2025rest};
\item Rest shape and material stiffness with penalty: simultaneous optimization of rest shape and material stiffness parameters with enforcement of $\constraint(\parameter) = 0$ via quadratic penalty \cite{NoceWrig06} (which is equivalent to ours with $\multiplier = 0$);
\item Rest shape and material stiffness with ALM (ours): rest shape and material stiffness parameter optimization with enforcement of $\constraint(\parameter) = 0$ as a hard constraint via ALM.
\end{enumerate}


\begin{figure}
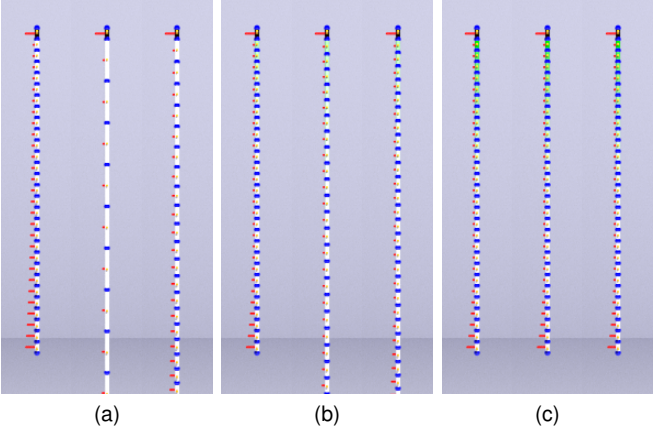
 
\centering
\subfloat[\label{1a}]{%
    \adjincludegraphics[trim={{0.45\width} {0.0\height} {0.45\width} {0.0\height}}, clip, width=0.105\linewidth]
    {figures/\figDir/rest_shape_vertical_rest_shape_0001.png}%
    \adjincludegraphics[trim={{0.45\width} {0.0\height} {0.45\width} {0.0\height}}, clip, width=0.105\linewidth]
    {figures/\figDir/rest_shape_vertical_rest_shape_0050.png}%
    \adjincludegraphics[trim={{0.45\width} {0.0\height} {0.45\width} {0.0\height}}, clip, width=0.105\linewidth]
    {figures/\figDir/rest_shape_vertical_rest_shape_0300.png}%
}
\ 
\subfloat[\label{1b}]{%
    \adjincludegraphics[trim={{0.45\width} {0.0\height} {0.45\width} {0.0\height}}, clip, width=0.105\linewidth]
    {figures/\figDir/rest_shape_vertical_penalty_0001.png}%
    \adjincludegraphics[trim={{0.45\width} {0.0\height} {0.45\width} {0.0\height}}, clip, width=0.105\linewidth]
    {figures/\figDir/rest_shape_vertical_penalty_0050.png}%
    \adjincludegraphics[trim={{0.45\width} {0.0\height} {0.45\width} {0.0\height}}, clip, width=0.105\linewidth]
    {figures/\figDir/rest_shape_vertical_penalty_0300.png}%
}
\
\subfloat[\label{1c}]{%
    \adjincludegraphics[trim={{0.45\width} {0.0\height} {0.45\width} {0.0\height}}, clip, width=0.105\linewidth]
    {figures/\figDir/rest_shape_vertical_alm_0001.png}%
    \adjincludegraphics[trim={{0.45\width} {0.0\height} {0.45\width} {0.0\height}}, clip, width=0.105\linewidth]
    {figures/\figDir/rest_shape_vertical_alm_0050.png}%
    \adjincludegraphics[trim={{0.45\width} {0.0\height} {0.45\width} {0.0\height}}, clip, width=0.105\linewidth]
    {figures/\figDir/rest_shape_vertical_alm_0300.png}%
}
\caption{
Evaluation with a vertical strand. The first edge (black) is fixed so its stretching stiffness parameter is undefined. The white/green edges represent lower/higher stiffness parameters. (a) Rest shape only. (b) Rest shape and material stiffness with penalty. (c) Rest shape and material stiffness with ALM (ours). With the rest shape only optimization, material stiffness parameters are unchanged (edges stay white), failing to achieve static equilibrium with rest shape parameters within their box constraints (a). Optimizing the material stiffness parameters additionally stiffens the edges and thus reduces the necessary rest shape changes. Treating the zero net force constraint as a hard constraint enables more significant stiffness changes to achieve static equilibrium (c), compared to using a soft one (b).
}
\label{fig:verticalStrand} 
\end{figure}

\subsubsection{Vertical Strand}
We test with a vertical strand discretized with 30 vertices (Figure \ref{fig:verticalStrand}) using $\restLength_{\mathrm{min}} = 10^{-2}$ and $\stretchCoefFinal = 10^{3}$.

Optimizing only rest shape parameters fails to achieve static equilibrium. Additionally optimizing stiffness parameters reduces the necessary rest shape changes, but enforcing $\constraint(\parameter) = 0$ as a soft constraint via a quadratic penalty also fails. This failure occurs because the penalty from material stiffness changes is large and comparable to the penalty from violation of $\constraint(\parameter) = 0$, leading to inadequate adjustment of the stiffness parameters. By contrast, our method guarantees perfect static equilibrium, as ALM enforces $\constraint(\parameter) = 0$ as a hard constraint. This allows for adequate adjustments to stiffness parameters, ensuring rest shape parameters remain within their box constraints while achieving zero net forces.

The rest shape only optimization takes 8 Newton iterations, 5.6 MPRGP iterations on average, and 8.1 ms for the entire optimization process, while the rest shape and material stiffness optimization with penalty takes 7 Newton iterations, 73.9 MPRGP iterations, and 19.5 ms for the optimization. The increased cost is primarily due to the larger system size ($7N-14$ for rest shape and material stiffness vs. $4N-8$ for rest shape only) with more non-zeros and increased MPRGP iterations (due to the increased complexity of the systems). Our method takes 25 Newton iterations, 66.0 MPRGP iterations, and 58.3 ms for the optimization. The increased cost compared to the approach with penalty is due to the updates of Lagrange multipliers, which modify the optimization problem. Consequently, our method requires around $3\times$ more Newton iterations and computational time. Although our method is approximately $7\times$ more costly than the rest shape only optimization \cite{Takahashi2025rest}, these methods are performed only once per scene as a preprocess (i.e., no additional cost during forward simulation). Thus, we believe that our approach, which guarantees static equilibrium with minimal stiffness changes while eliminating tedious manual stiffness tuning, is a practical and attractive alternative.


\begin{figure}
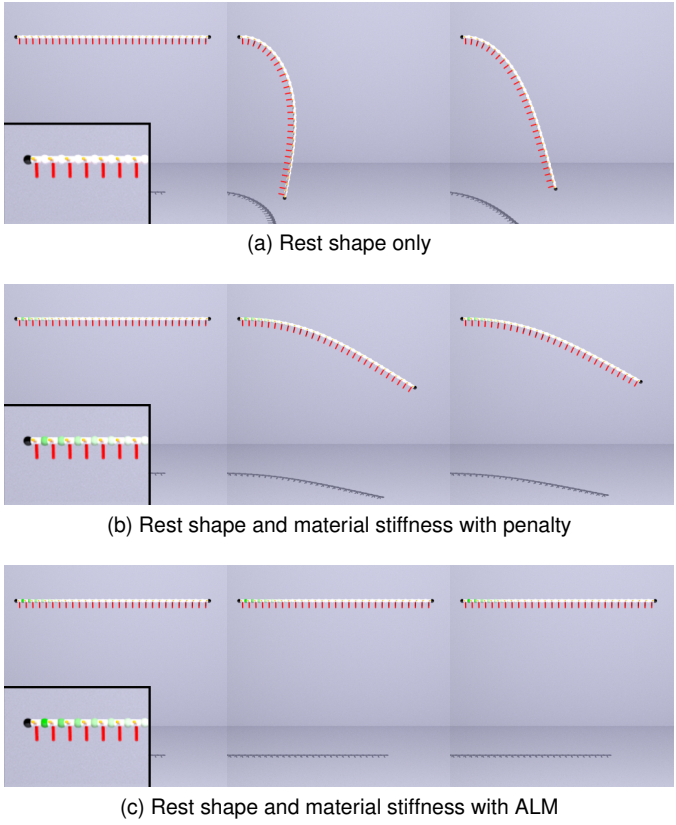
 
\centering
\subfloat[Rest shape only\label{1a}]{%
    \adjincludegraphics[trim={{0.23\width} {0.0\height} {0.23\width} {0.05\height}}, clip, width=0.333\linewidth]
    {figures/\figDir/rest_shape_horizontal_rest_shape_0001.png}%
    \adjincludegraphics[trim={{0.23\width} {0.0\height} {0.23\width} {0.05\height}}, clip, width=0.333\linewidth]
    {figures/\figDir/rest_shape_horizontal_rest_shape_0060.png}%
    \adjincludegraphics[trim={{0.23\width} {0.0\height} {0.23\width} {0.05\height}}, clip, width=0.333\linewidth]
    {figures/\figDir/rest_shape_horizontal_rest_shape_0300.png}%
}%
\\
\subfloat[Rest shape and material stiffness with penalty\label{1a}]{%
    \adjincludegraphics[trim={{0.23\width} {0.0\height} {0.23\width} {0.05\height}}, clip, width=0.333\linewidth]
    {figures/\figDir/rest_shape_horizontal_penalty_0001.png}%
    \adjincludegraphics[trim={{0.23\width} {0.0\height} {0.23\width} {0.05\height}}, clip, width=0.333\linewidth]
    {figures/\figDir/rest_shape_horizontal_penalty_0060.png}%
    \adjincludegraphics[trim={{0.23\width} {0.0\height} {0.23\width} {0.05\height}}, clip, width=0.333\linewidth]
    {figures/\figDir/rest_shape_horizontal_penalty_0300.png}%
}%
\\
\subfloat[Rest shape and material stiffness with ALM\label{1a}]{%
    \adjincludegraphics[trim={{0.23\width} {0.0\height} {0.23\width} {0.05\height}}, clip, width=0.333\linewidth]
    {figures/\figDir/rest_shape_horizontal_alm_0001.png}%
    \adjincludegraphics[trim={{0.23\width} {0.0\height} {0.23\width} {0.05\height}}, clip, width=0.333\linewidth]
    {figures/\figDir/rest_shape_horizontal_alm_0060.png}%
    \adjincludegraphics[trim={{0.23\width} {0.0\height} {0.23\width} {0.05\height}}, clip, width=0.333\linewidth]
    {figures/\figDir/rest_shape_horizontal_alm_0300.png}%
}%
\caption{
Evaluation with a horizontal strand. Material stiffness parameters for bending are undefined for the first and last vertices (black). The white and green vertices represent lower and higher stiffness parameters, respectively. The insets provide enlarged views for clarity. The rest shape only optimization fails to achieve static equilibrium (a). While simultaneously optimizing rest shape and material stiffness parameters better enforces zero forces, treating the zero net force constraint as a soft constraint still fails (b), but as a hard constraint it succeeds in achieving the horizontal static equilibrium (c).
}
\label{fig:horizontalStrand} 
\end{figure}

\subsubsection{Horizontal Strand}
We experiment with a horizontal strand discretized with 30 vertices (Figure \ref{fig:horizontalStrand}) using $\curvRange = 0.2$. Similar to the vertical strand case, optimizing only the rest shape parameters fails to achieve static equilibrium (despite the relatively stiff strand with $\bendCoefFinal = 10^{9}$) due to tightly bounded rest curvature changes. While simultaneous optimization of the rest shape and material stiffness aims to achieve zero net forces, it still fails when enforcing $\constraint(\parameter) = 0$ as a soft constraint using the penalty method. By contrast, our method successfully attains static equilibrium. The computational costs follow a trend similar to that observed with the vertical strand. The optimization takes 7.5 ms for the rest shape only, 17.9 ms for the rest shape and material stiffness with penalty, and 47.7 ms for our method.


\subsection{Box-Constrained Quadratic Program Solver Comparisons}
We compare our method with various schemes to solve the BCQP subproblem \eqref{eq:bcqp} using the scene shown in Figure \ref{fig:activeset}.


\begin{figure}[tb]
\captionsetup[subfigure]{aboveskip=0mm,belowskip=0mm}
\centering
\adjincludegraphics[trim={{0.0\width} {0.0\height} {0.0\width} {0.\height}}, clip, width=1.0\linewidth]
{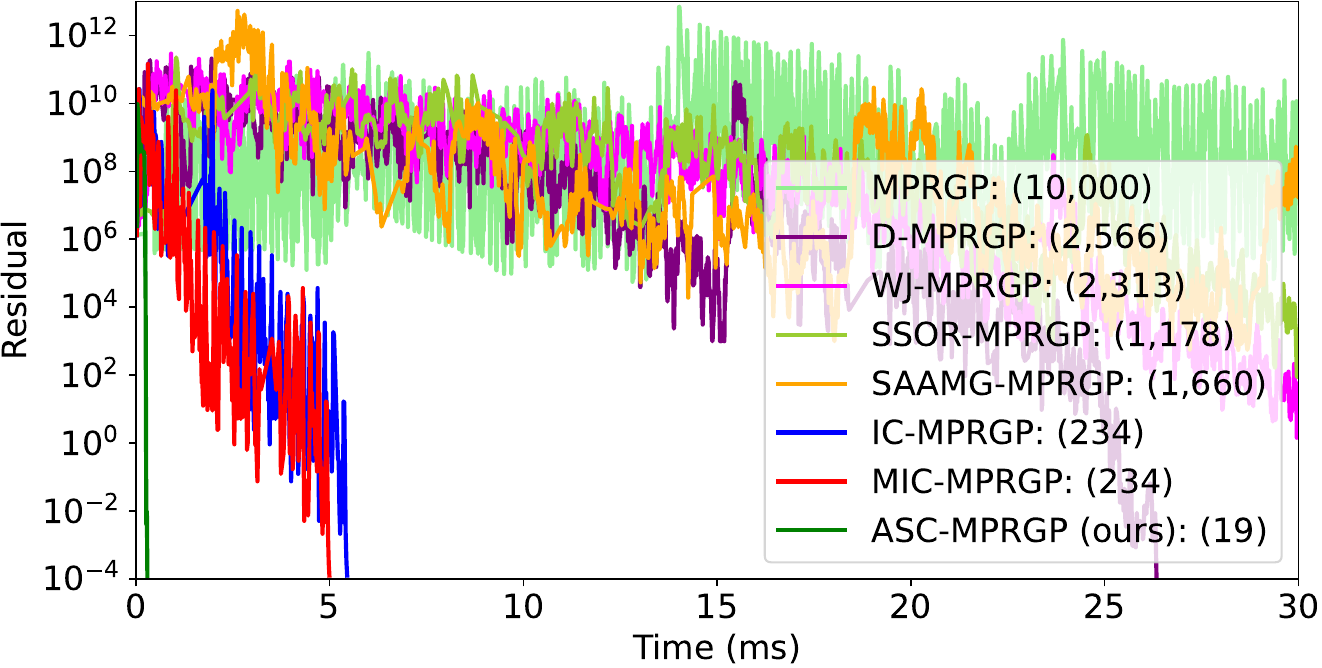}
\caption{
Log-scale profiles of residuals over time with various preconditioning techniques in MPRGP. The numbers in the last parentheses represent MPRGP iteration counts. Our ASC-MPRGP converges $18\times$ faster than IC-MPRGP.
\vspace{-3mm}
}
\label{fig:mprgp}
\end{figure}

\subsubsection{MPRGP Preconditioners}
We first examine our method against various preconditioning techniques for MPRGP. Specifically, we compare the following schemes:
\begin{enumerate}
\item MPRGP: MPRGP without preconditioning \cite{Dostal2005};
\item D-MPRGP: MPRGP with diagonal preconditioning;
\item WJ-MPRGP: MPRGP with two weighted Jacobi iterations with a weighting factor $\omega = 0.5$;
\item SSOR-MPRGP: MPRGP with symmetric successive-over-relaxation (SSOR) preconditioning (serial forward and backward passes) with a weighting factor $\omega = 1.2$;
\item SAAMG-MPRGP: MPRGP with SAAMG preconditioning \cite{Takahashi2023};
\item IC-MPRGP: MPRGP with IC preconditioning \cite{Narain:2010:FGM:1882261.1866195};
\item MIC-MPRGP: MPRGP with MIC preconditioning \cite{Narain:2010:FGM:1882261.1866195};
\item ASC-MPRGP (ours): MPRGP with our ASC preconditioning. 
\end{enumerate}

Figure \ref{fig:mprgp} shows log-scale profiles of convergence over time. Due to the stiffness of our system, MPRGP without preconditioning failed to converge within 10,000 iterations and was terminated. While WJ-MPRGP and SSOR-MPRGP effectively reduce the number of MPRGP iterations required, their preconditioning costs are non-negligible, and D-MPRGP performs slightly faster. Although SAAMG-MPRGP can be effective for systems with an M-matrix \cite{Takahashi2023}, it underperforms on our non-M-matrix system. By contrast, Cholesky-based preconditioners are particularly effective because they exploit the banded structure of the system with minimal overhead for a single factorization per MPRGP solve. IC-MPRGP requires significantly fewer iterations than D-MPRGP, and its moderate preconditioning cost results in much faster overall performance. Given the nearly fully filled band, MIC-MPRGP performs almost identically to IC-MPRGP. Our ASC-MPRGP, which employs full Cholesky factorization, benefits from the complete Cholesky factors without the limitations of (M)IC-MPRGP, which uses incomplete Cholesky with a GS fallback to avoid breakdown \cite{Bridson2007course}. Consequently, while IC-MPRGP converges in 234 iterations taking 5.4 ms, our ASC-MPRGP converges in just 19 iterations taking 0.3 ms, achieving around $18\times$ faster performance.


\begin{figure}[tb]
\captionsetup[subfigure]{aboveskip=0mm,belowskip=0mm}
\centering
\adjincludegraphics[trim={{0.0\width} {0.0\height} {0.0\width} {0.\height}}, clip, width=1.0\linewidth]
{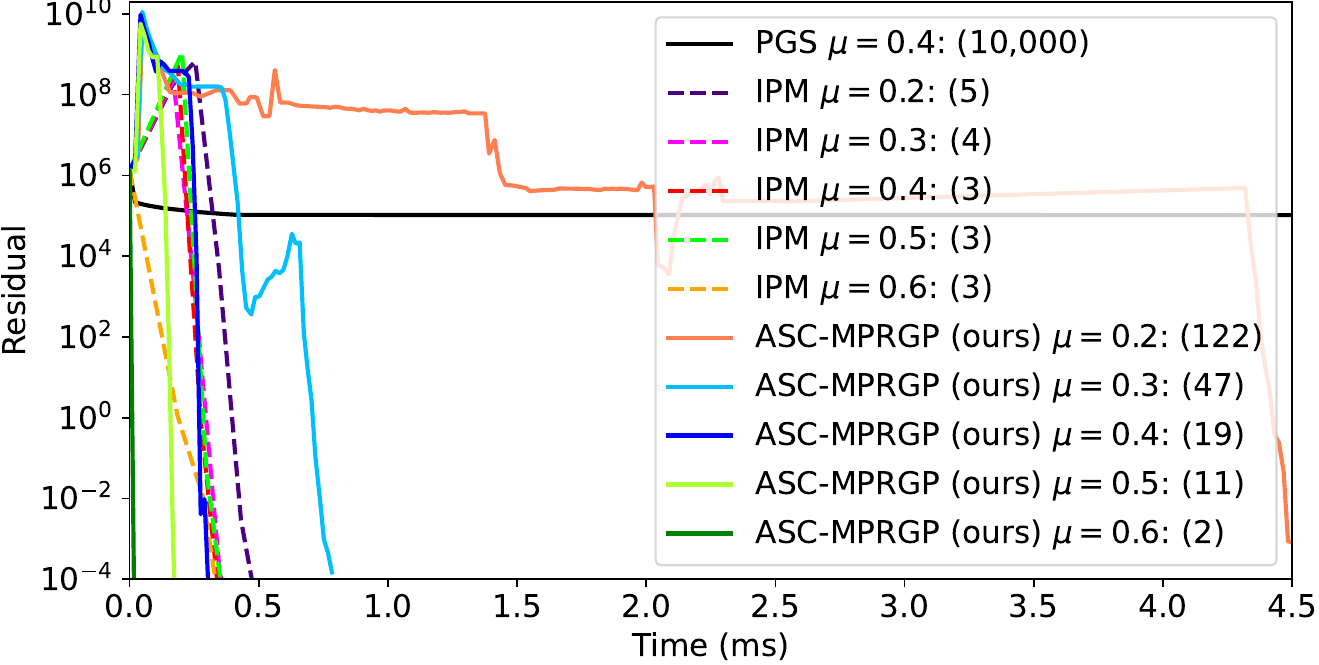}
\caption{
Log-scale profiles of residuals over time with various BCQP solvers and permitted curvature change $\curvRange$. The numbers in the last parentheses indicate solver iteration counts. Our ASC-MPRGP outperforms IPM for $\curvRange \geq 0.4$.
}
\label{fig:bcqp}
\end{figure}

\begin{table}[tb]
\centering
\caption{
Performance numbers for IPM and our ASC-MPRGP with various $\curvRange$. Timing is given in milliseconds, and the numbers in the parentheses represent the iteration counts.
}
\scalebox{0.86}{
\begin{tabular}{c|rrrrrrrrrr}\hline
Scheme \textbackslash\ $\curvRange$ & 0.2 & 0.3 & 0.4 & 0.5 & 0.6\\
\hline \hline
IPM       & 0.517 (5) & 0.434 (4) & 0.361 (3) & 0.388 (3) & 0.375 (3)\\
ASC-MPRGP & 4.518 (122) & 0.783 (47) & 0.305 (19) & 0.184 (11) & 0.039 (2)\\
\hline
\end{tabular}}
\label{tab:bcqp}
\end{table}

\subsubsection{BCQP Solvers}
Next, we compare our method with other BCQP solvers. We consider the following schemes:
\begin{enumerate}
\item PGS: projected GS;
\item IPM: interior point method \cite{Takahashi2024qp} with a Cholesky-based direct solver for fully unconstrained inner linear systems;
\item ASC-MPRGP (ours).
\end{enumerate}
In this comparison, we exclude the penalty-based approach as it cannot accurately enforce the box constraints (comparisons between the penalty-based and active-set approaches are given in Sec. \ref{sec:penalty_active_set}). Figure \ref{fig:bcqp} shows log-scale profiles of convergence over time, and Table \ref{tab:bcqp} summarizes the performance numbers (excluding PGS).

Although PGS should be able to strictly enforce the box constraints, it was slow to converge and terminated after 10,000 iterations. While IPM \cite{Takahashi2024qp} can converge with a small number of iterations for all values of $\curvRange$, each iteration is costly because IPM updates the system diagonals, necessitating a full Cholesky factorization for each updated system. By contrast, our ASC-MPRGP requires no system updates and enables the reuse of the Cholesky factor by combining it with active sets which may change in each MPRGP iteration. Thus, per-iteration cost is relatively small, enabling ASC-MPRGP to outperform IPM \cite{Takahashi2024qp} with approximately $\curvRange \geq 0.4$ (our default is $\curvRange = 1$). While our method can be slower than IPM when $\curvRange$ is small (because MPRGP needs more iterations to update active sets and does not fully leverage our preconditioner), we believe that such cases are infrequent in practical applications.


\begin{figure*}
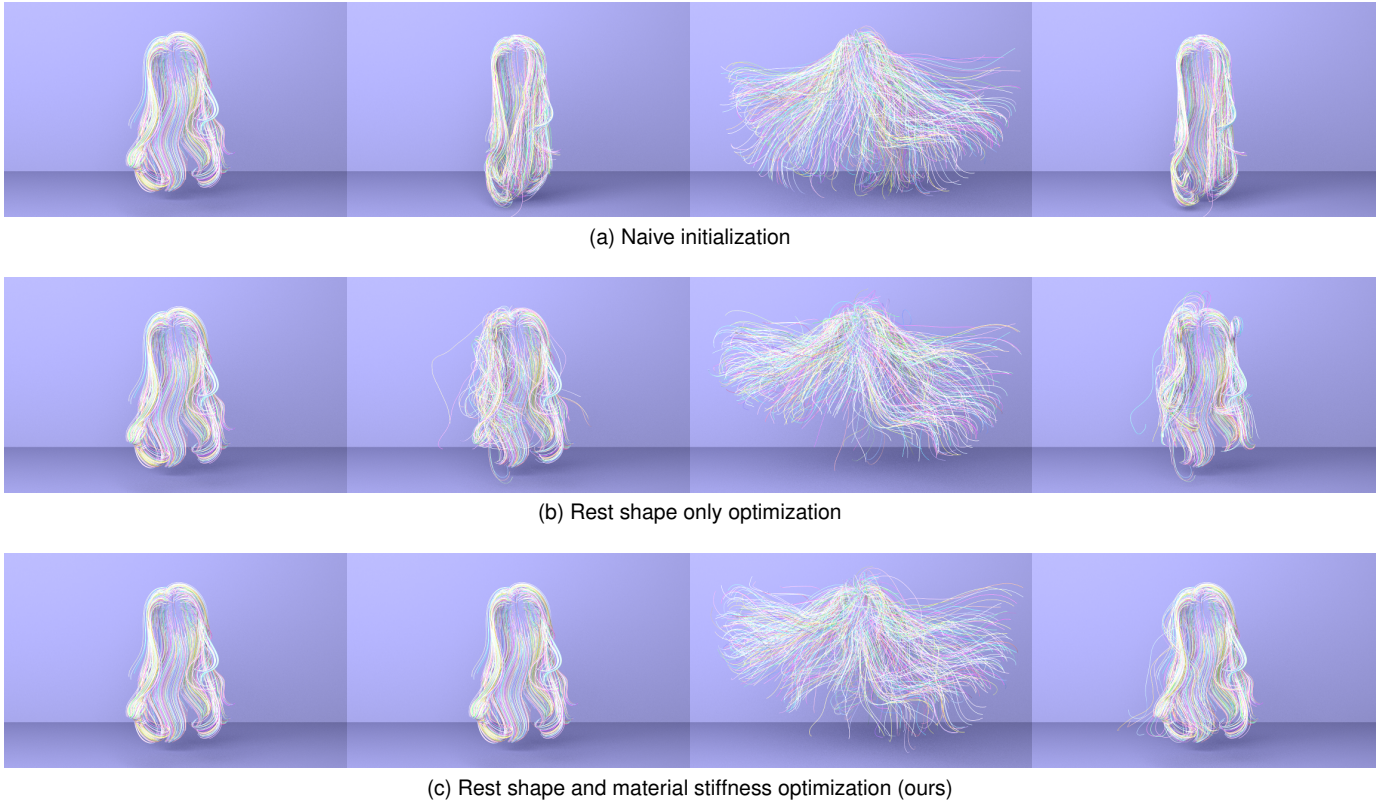
 
\centering
\subfloat[Naive initialization\label{1a}]{%
    \adjincludegraphics[trim={{0.05\width} {0.0\height} {0.05\width} {0.0\height}}, clip, width=0.25\linewidth]
    {figures/\figDir/hair_naive_initialization_0001.png}%
    \adjincludegraphics[trim={{0.05\width} {0.0\height} {0.05\width} {0.0\height}}, clip, width=0.25\linewidth]
    {figures/\figDir/hair_naive_initialization_0181.png}%
    \adjincludegraphics[trim={{0.05\width} {0.0\height} {0.05\width} {0.0\height}}, clip, width=0.25\linewidth]
    {figures/\figDir/hair_naive_initialization_0310.png}%
    \adjincludegraphics[trim={{0.05\width} {0.0\height} {0.05\width} {0.0\height}}, clip, width=0.25\linewidth]
    {figures/\figDir/hair_naive_initialization_1200.png}%
}%
\\
\subfloat[Rest shape only optimization\label{1a}]{%
    \adjincludegraphics[trim={{0.05\width} {0.0\height} {0.05\width} {0.0\height}}, clip, width=0.25\linewidth]
    {figures/\figDir/hair_rest_shape_only_0001.png}%
    \adjincludegraphics[trim={{0.05\width} {0.0\height} {0.05\width} {0.0\height}}, clip, width=0.25\linewidth]
    {figures/\figDir/hair_rest_shape_only_0181.png}%
    \adjincludegraphics[trim={{0.05\width} {0.0\height} {0.05\width} {0.0\height}}, clip, width=0.25\linewidth]
    {figures/\figDir/hair_rest_shape_only_0310.png}%
    \adjincludegraphics[trim={{0.05\width} {0.0\height} {0.05\width} {0.0\height}}, clip, width=0.25\linewidth]
    {figures/\figDir/hair_rest_shape_only_1200.png}%
}%
\\
\subfloat[Rest shape and material stiffness optimization (ours)\label{1a}]{%
    \adjincludegraphics[trim={{0.05\width} {0.0\height} {0.05\width} {0.0\height}}, clip, width=0.25\linewidth]
    {figures/\figDir/hair_ours_0001.png}%
    \adjincludegraphics[trim={{0.05\width} {0.0\height} {0.05\width} {0.0\height}}, clip, width=0.25\linewidth]
    {figures/\figDir/hair_ours_0181.png}%
    \adjincludegraphics[trim={{0.05\width} {0.0\height} {0.05\width} {0.0\height}}, clip, width=0.25\linewidth]
    {figures/\figDir/hair_ours_0310.png}%
    \adjincludegraphics[trim={{0.05\width} {0.0\height} {0.05\width} {0.0\height}}, clip, width=0.25\linewidth]
    {figures/\figDir/hair_ours_1200.png}%
}%
\caption{
Hair simulation with complex strand geometry. Hair strands significantly sag with naive initialization. Rest shape only optimization suffers from some sagging and unnatural hair lifting. Our method successfully achieves static equilibrium, exhibits natural motions driven by the prescribed movements of the root vertices, and then restores the hairstyle to a form closely resembling the original.
\vspace{-3mm}
}
\label{fig:hairs} 
\end{figure*}

\subsection{Complex Strand Geometry}
To evaluate the efficacy of our method with complex strand geometry, we experiment with hair data publicly released by Hu et al. \cite{Hu2015hair}. Figure \ref{fig:hairs} compares our method with naive initialization and rest shape only optimization, using an asset with 915 strands (each is discretized with 100 vertices). In this comparison, we use an extra soft material with $\stretchCoefFinal = \bendCoefFinal = \twistCoefFinal = 10^{8}$ to clearly demonstrate differences, and $\curvRange = 0.5$.

With naive initialization, hair strands sag significantly due to gravity at the start of the simulation, eventually settling into sagged states after some root vertex movement. With rest shape only optimization, some strands still suffer from sagging since the allowed rest shape changes are limited by the box constraints. Conversely, other strands experience unnatural lifting because this approach attempts to achieve static equilibrium solely through adjustments to the rest shape parameters, yielding significant rest shape changes (even with the box constraints) and thus continuously unstable strand configurations that do not settle. Hence the rest shape only optimization fails both to achieve static equilibrium (with insufficient rest shape changes) and to ensure stable configurations (due to the excessive rest shape changes), indicating that the range parameter for box constraints $\curvRange$ is simultaneously too low for some strands and too high for others; therefore, merely adjusting $\curvRange$ cannot always fix these failures. By contrast, our method enables strands to achieve the perfect static equilibrium even with  complex hair geometry through the smaller rest shape changes facilitated by simultaneous optimization of material stiffness. It achieves plausible motions and allows the strands to gradually settle and return towards the original hair styles, while naive initialization and rest shape only optimization do not.

Rest shape only optimization used 2.0 Newton iterations (with frequent failures in backtracking line search, leading to early termination) and 2.5 MPRGP iterations per solve on average (occasionally terminating at 100 iterations due to convergence issues), taking 0.6 s for the entire optimization process, whereas our method used 148.8 Newton iterations and 1.2 MPRGP iterations, taking 84.4 s. Although our method incurs higher costs than the rest shape only optimization (which fails to converge, thus suffering from sagging, unnatural lifting, and stability problems), we believe that the achieved static equilibrium justifies this one time initialization cost for forward simulations (which take 2.0 s per frame).



\section{Conclusions and Future Work}
We have proposed our parameter optimization framework that ensures static equilibrium of DER-based strands. Additionally, we presented an active-set Cholesky preconditioner to accelerate the convergence of MPRGP. We demonstrated the efficacy of our method in a wide range of examples. Below, we discuss tradeoffs inherent to our method and promising research directions for future work.


\subsection{Undesirable Local Minima}
While our method ensures that strands achieve static equilibrium, certain perturbations may cause them to fall into other local minima in forward simulations, which may not align with user expectations. Although our solver enforces rest shape changes within specified box constraints to mitigate the risk of encountering such local minima, determining optimal values for these constraints can be challenging. The desirability of the resulting behaviors is often subjective, and the optimal values may vary depending on strand geometry and material stiffness. To accommodate diverse scenarios, it could be advantageous to assign different values of $\curvRange$ for each strand.

In contrast to optimization methods focused solely on the rest shape parameters \cite{Takahashi2025rest}, our approach allows for modifications to the material stiffness, which can be perceived as undesirable. While $\Wmat$ in \eqref{eq:parameter_optimization} assists in balancing changes between the rest shape and material stiffness parameters, a potentially more effective strategy may involve conducting separate and iterative parameter optimizations for the rest shape and material stiffness, similar to block coordinate descent. Furthermore, to ensure smoothly varying material stiffness, it may be beneficial to employ a Laplacian-based regularizer or, alternatively, to optimize a limited set of representative stiffness parameters, which has the added advantage of reducing memory usage and optimization costs.


\subsection{Active-Set Cholesky Preconditioner}
As our ASC preconditioner is based on a Cholesky-based direct solver which is particularly effective for stiff SPD linear systems, it is worth evaluating our preconditioner with stiff BCQPs. In particular, our preconditioner should work with tree structures which form block-structured systems (instead of banded ones) without introducing any fill-in at off-diagonal blocks between the block matrices \cite{Herholz2018}. It also seems promising to extend our preconditioner to (incomplete) LDLT factorization \cite{Davis2016direct,Greif2017ldlt} and to explore symmetric indefinite solvers that can handle box constraints. With small $\curvRange$, many box constraints can be activated, necessitating additional MPRGP iterations to manage active sets while failing to fully leverage our ASC preconditioning. In such cases, exploring active-set-free approaches, such as interior point methods \cite{NoceWrig06,Takahashi2024qp}, may be promising. Although parallel execution over strands rendered extensive parallelization unnecessary in our framework, it would be valuable to optimize our preconditioner for parallel execution to fully utilize many-core architectures, particularly for strands discretized with a large number of vertices.


\subsection{More General Inverse Problems}
Supporting anisotropic and inhomogeneous strands, two-end clamping, and frictional contacts would be useful. In particular, incorporating contact forces in the parameter optimization would prevent extra stiffening and rest shape changes, giving more plausible results (e.g., for hairs on a head) \cite{Derouet-Jourdan:2013:IDH:2508363.2508398,Hsu2022sag,Hsu2023sag}. To predict strand dynamics over time, developing differentiable physics approaches, such as those employing the adjoint method \cite{Perez2015rod,Panetta2019}, appears promising \cite{chen2024differentiablediscreteelasticrods}. Extending our method to applications involving 2D/3D structured materials would also be of interest. While our method is guaranteed to reach static equilibrium at solver convergence, convergence conditions may depend on strand geometry and properties; thus, tuning solver parameters may be necessary in challenging scenarios. Additionally, solving the optimization problem without primal-dual decoupling could enhance the likelihood of convergence.

\vspace{-3mm}
\section*{Acknowledgments}
We thank the anonymous reviewers for their valuable suggestions, and Bo Yang, Qingyuan Zheng, Rundong Wu, and Yu Ju Chen for early discussions. We acknowledge the authors of \cite{Hu2015hair} for publicly releasing the hair dataset. This work was supported in part by the Natural Sciences and Engineering Research Council of Canada (Grant RGPIN-2021-02524).
\vspace{-3mm}
\bibliographystyle{IEEEtran}
\bibliography{source/reference}

\vspace{-12mm}
\begin{IEEEbiography}[{\vspace{-5mm}\includegraphics[width=1in,height=1in,clip,keepaspectratio]{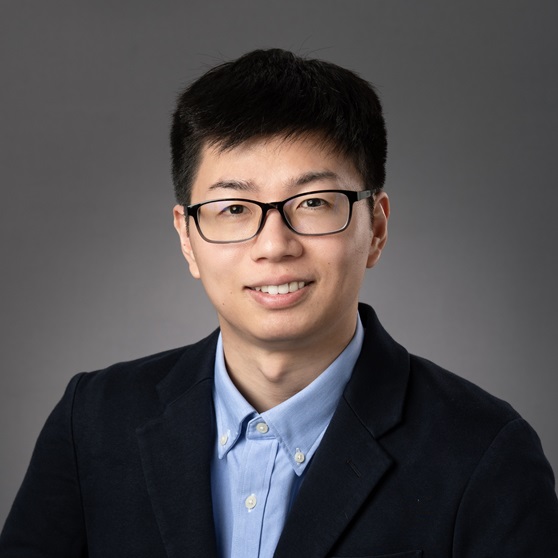}}]{Tetsuya Takahashi}
is a Senior Researcher at Tencent America. He was a Software Engineer at Adobe. He earned his M.S. and Ph.D. from the University of North Carolina at Chapel Hill in 2017 and 2020, respectively, and B.S. and M.S. from Keio University in 2012 and 2014, respectively. His research interests include physically-based simulation, numerical optimization, and geometry processing.
\end{IEEEbiography}
\vspace{-9mm}
\begin{IEEEbiography}
[{\vspace{-2mm}\includegraphics[width=1in,height=1.25in,clip,keepaspectratio]{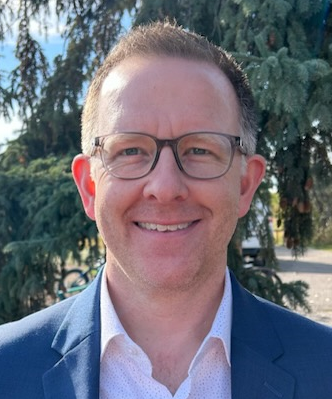}}]{Christopher Batty}
is an Associate Professor in the David R. Cheriton School of Computer Science at the University of Waterloo in Ontario, Canada. He received his PhD from the University of British Columbia in 2010 and was a Banting Postdoctoral Fellow at Columbia University from 2011 to 2013. His research is primarily focused on the development of novel physical simulation techniques for applications in computer graphics and computational physics, with an emphasis on the diverse behaviors of fluids.
\end{IEEEbiography}



\end{document}